\documentclass[10pt,journal]{IEEEtran}

\usepackage[T1]{fontenc}
\usepackage[utf8]{inputenc}

\usepackage[normalem]{ulem}
\usepackage{pifont}
\usepackage{amsmath,amssymb}
\usepackage{siunitx}
\usepackage{multirow}
\usepackage{booktabs}
\usepackage[caption=false, labelformat=empty]{subfig}
\usepackage[export]{adjustbox}
\usepackage{textcomp}
\usepackage{stfloats}
\usepackage{graphicx}
\usepackage{xcolor}
\usepackage{cite}
\usepackage[draft=false, hyperindex, plainpages=false, pdfpagelabels=true, breaklinks=true, bookmarksnumbered=true]{hyperref}

\newcommand{\etal}{\textit{et~al.}}%
\newcommand{\cmark}{\ding{51}}%
\newcommand{\xmark}{\ding{55}}%

\newcommand{\reviewed}[1]{#1}%

\begin{document}

\title{Combining Pre- and Post-Demosaicking Noise Removal for RAW Video}

\author{M. Sánchez-Beeckman, A. Buades, N. Brandonisio, B. Kanoun
	\thanks{M. Sánchez-Beeckman and A. Buades are with Institute of Applied Computing and Community Code (IAC3), Parc BIT, C/ Blaise Pascal 7 and with the Dept.\ of Mathematics and Computer Science, Universitat de les Illes Balears, Cra.~de Valldemossa km.~7.5, E-07122 Palma, Spain.}
	\thanks{N. Brandonisio and B. Kanoun are with Huawei Technologies France.}
	\thanks{This publication is part of the Grant PID2021-125711OB-I00 funded by MCIN/AEI/ 10.13039/501100011033 and the European Union NextGeneration EU/PRTR.}
	\thanks{M. Sánchez-Beeckman is also supported by the Conselleria de Fons Europeus, Universitat i Cultura del Govern de les Illes Balears through grant FPU2023-011-C.}
}

\markboth{}%
{Sánchez-Beeckman \MakeLowercase{\textit{et al.}}: Combining Pre- and Post-Demosaicking Noise Removal for RAW Video}


\maketitle

\begin{abstract}
	Denoising is one of the fundamental steps of the processing pipeline that converts data captured by a camera sensor into a display-ready image or video.
\reviewed{It is generally performed early in the pipeline, usually before demosaicking, although studies swapping their order or even conducting them jointly have been proposed.}
\reviewed{With the advent of deep learning, the quality of denoising algorithms has steadily increased.}
	Even so, modern neural networks still have a hard time adapting to new noise levels and scenes, which is indispensable for real-world applications.
    \reviewed{With those in mind, we propose} a self-similarity-based denoising scheme that weights both a pre- and a post-demosaicking denoiser \reviewed{for Bayer-patterned CFA video data}.
	We show that a balance between the two leads to better image quality, and we empirically find that higher noise levels benefit from a higher influence pre-demosaicking.
	We also integrate temporal trajectory prefiltering steps before each denoiser, which further improve texture reconstruction.
	The proposed method only requires an estimation of the noise model at the sensor, accurately adapts to any noise level, and is competitive with the state of the art, making it suitable for real-world videography.
\end{abstract}


\section{Introduction}
\IEEEPARstart{D}{igital} photographs and videos, as seen on a screen, are the end result of a complex chain of transformations that process the energy of incident photons in an observed scene into pixel values~\cite{Ramanath2005ProcessingPipeline}.
The goal of these transformations---henceforth referred to as the \reviewed{Image Signal Processing (ISP), or simply the processing pipeline}---is to ensure that the original visual information in that scene is faithfully preserved.

Digital cameras capture images using an electronic image sensor, most commonly \reviewed{Charge Coupled Device} (CCD) or \reviewed{Complementary Metal Oxide Semiconductor} (CMOS), which is a device that converts light into an electrical signal.
Light that is captured by the sensor passes through a Color Filter Array (CFA), a patterned layer that allows only a specific wavelength---red, green or blue---to pass through at each pixel.
Usually, the CFA follows a \(2\times2\) Bayer pattern, where the green filters are duplicated to mimic the human eye's higher sensitivity to green light~\cite{Palum2001Bayer}.
The values that are read at this stage are called the RAW data, and they are a direct representation of the number of photons that were captured by each pixel~\cite{ElGamal2005CMOS}.
The remaining color information must be interpolated---demosaicked---from the surrounding pixels in order to create a full-color image.
In addition, other transformations are typically applied during the processing pipeline to reconstruct the captured scene: color corrections such as white balance, gamma correction and tone mapping, as well as other adjustments to compensate for lens shading and to sharpen the~results.

Physical factors such as sensor size, temperature and electronic interference, as well as the random nature of the photoelectric effect and the conversion from electrical signals to digital data make the presence of noise during image acquisition inevitable~\cite{Bigas2006CMOS}.
As noise is usually undesirable, especially when its amplitude obfuscates the details of the scene, denoising is an indispensable step of the ISP.

Noise is much harder to remove after an image or video has undergone the complete processing pipeline.
The pixel values of a processed photograph are usually compressed, they have a lower bit depth than the original RAW data, they no longer hold a direct relation with the number of photons, and their noise has become heavily correlated~\cite{Park2009DenoisingDemosaicking}.
The limitations of denoising after the full processing pipeline are especially pronounced in low-light conditions, where the signal is weaker and noise tends to be more noticeable~\cite{Wei2020Physics}.
Darker scenes usually need to be captured with higher camera ISO \reviewed{sensitivity} values to be visible, at the expense of amplifying the noise even further.
Longer exposure times can be used to mitigate this effect, but they can also introduce motion blur, being an unrealistic solution in the case of video.

These issues have led to the common practice of performing denoising in the RAW domain before demosaicking, where the noise is spatially uncorrelated and the pixel values are still linearly related to the number of photons~\cite{Kalevo2002NoiseReduction,Park2009DenoisingDemosaicking,Lee2017Denoising}.
Despite this, Jin~\etal~\cite{Jin2020Dilemma} have provided \reviewed{empirical results showing} that denoising just after demosaicking may lead to a better texture recovery, on the condition that a stronger filtering is applied.
Some works have followed that observation and proposed denoising methods in the demosaicked domain~\cite{Ehret2019F2F,Dewil2021MF2F}.
The same authors of~\cite{Jin2020Dilemma} recently argued in~\cite{Guo2023BestCombine} that, for high noise levels, there is a moderate benefit in applying a partial RAW denoising followed by demosaicking and a second denoising on the RGB image.
\reviewed{The study is conducted using black box optimization with two BM3D~\cite{Dabov2007BM3D} denoisers, and applies to synthetic Bayer CFA single image data with added uniform white Gaussian noise. The authors conclude that such improvement is too marginal to justify the use of a combined approach and that in practice those noise levels do not appear in real RAW data.}
In the case of RAW video denoising, the state of the art is still greatly composed of methods that denoise before demosaicking~\cite{Yue2020RViDeNet,Paliwal2021MaskDnGAN,Sheth2021UDVD,Li2022Multifrequency,Ostrowski2022BPEVD,Maggioni2021EMVD,Fu2022LLRVD,Li2022FloRNN}.
\reviewed{Joint denoising and demosaicking (JDD) schemes, despite trending for single images~\cite{Xing2021EndtoEnd,Guo2021JDD,Chen2021JDD,Zhang2022JDD,Guo2023Joint,Liu2020JDD,Li2024JDD}, can only be found sparsely in the video literature \cite{Jiang2019SMOID,Dewil2023RVDD}.}

\reviewed{Nowadays, research on denoising seems to have mostly moved on to the field of deep learning.}
Yet, RAW video denoising in this age is almost synonymous with mobile photography, which has its own limitations.
Mobile devices are constrained by their computational power and memory capacity, and they must work within strict bounds of power consumption and heat dissipation~\cite{Kamenicky2019Portable}.
These limitations are further aggravated when frames must be processed in a reasonable time.
Under those conditions, mobile devices may not be able to run complex deep learning models such as those that are used in the literature.
In this context, we set our objective on exploring how to improve classical video denoising algorithms so as to make them a viable alternative to neural networks for general video denoising tasks.

The main contribution of our work is the design of a scheme which balances the application of a classical self-similarity-based denoising algorithm pre- and post-demosaicking \reviewed{for Bayer-patterned RAW video sequences with real sensor noise.}
\reviewed{
    This solution is motivated by the fact that the CFA is highly aliased, which difficults patch comparison and detail recovery, often leading to unwanted checkerboard artifacts. 
    On the opposite end, the demosaicking process correlates the noise both spatially and in color.
    Such correlated noise is very difficult to remove compared to the (aproximately) Poisson-Gaussian noise in the CFA, especially for dark and indoor scenes that require a high ISO sensitivity.
    This leads to surmise that for good lighting conditions with a low ISO value (i.e.\ low noise variances), denoising should concentrate on the sequence after demosaicking in order to avoid aliasing issues.
    Conversely, when dealing with very noisy dark scenes and high ISO settings, noise should be reduced at the RAW level in order to mitigate the effects of significant noise correlation.
    These observations make apparent that the most straightforward and coherent configuration is a combination of two denoising steps, where the degree of filtering in the RAW domain is controlled by a parameter that depends on the ISO level.
    In fact, we find that, granting even a relatively small influence to one of the two denoising stages, this configuration yields meaningful improvements across all ISO levels, which differs from the conclusions drawn by~\cite{Guo2023BestCombine}.
}

\reviewed{
    With the proposed chain, we demonstrate that classical denoising algorithms can still produce results with a quality on par or even better than state-of-the-art deep learning-based models.
    In order to do so, we show that further leveraging temporal coherence is crucial for a correct texture reconstruction.
    Notably, we study the effects of temporal prefiltering on the quality of the recovered textures after denoising when applied before and after demosaicking.
    Lastly, we show that the proposed method shines when it comes to its adaptability to different kinds of sequences and noise levels, which are still a challenge for modern neural networks.
}

\section{Related Work}

\subsection{\reviewed{sRGB Video Denoising}}

Patch-based schemes such as NLMeans~\cite{Buades2005NLMeans}, BM3D~\cite{Dabov2007BM3D} and NLBayes~\cite{Lebrun2013NLBayes} make up the core of what we now consider the reference classical methods for image denoising.
Their nature makes them straightforward to adapt to image sequences, which can be done by simply extending the neighbouring area they use to adjacent frames.
In that fashion, NLMeans has been extended by Boulanger~\etal~\cite{Boulanger2007SpaceTime}; BM3D, by Dabov~\etal~\cite{Dabov2007VBM3D} into VBM3D\@; and NLBayes, by Arias and Morel~\cite{Arias2017Bayesian}.

Nevertheless, simply extending the search area to adjacent frames is not enough to fully exploit the temporal redundancy of video sequences.
A more sophisticated approach is to use motion compensation to help select coherent patches.
Yet another BM3D extension, VBM4D~\cite{Maggioni2011VBM4D}, uses motion vectors to build spatio-temporal volumes, which are stacked and denoised through collaborative filtering.
A similar idea to VBM4D is proposed in~\cite{Buades2016Patch}, where the authors apply a patch-based denoising method to a temporal window of frames warped with optical flow.
Robustness to motion estimation errors is attained by performing spatio-temporal patch comparisons with an occlusion mask.
That same strategy is adapted to signal dependent noise in~\cite{Buades2019Enhancement} by adding a noise estimation step, variance stabilization and a multiscale filtering scheme.

\reviewed{Deep learning models have quickly taken over as the preferred approach to single image denoising in the literature.}
Contrary to classical methods, however, transitioning to video has shown to be more complex, leading to a rich variety of architectures to take advantage of temporal redundancy.
Chen~\etal~\cite{Chen2016RNN} use a recurrent network architecture to learn to exploit latent spatio-temporal information.
Davy~\etal~\cite{Davy2019VNLnet} incorporate the classical prior of non-local self-similarity into a \reviewed{Convolutional Neural Network} (CNN) that predicts a clean image from spatio-temporal stacks of similar patches.
Relatedly, Vaksman~\etal~\cite{Vaksman2021PatchCraft} also exploit non-local self-similarity and temporal redundancy by tiling matched patches which are fed to a CNN\@.
Tassano~\etal~\cite{Tassano2019DVDnet} propose DVDnet, a CNN that uses separate spatial and temporal filtering blocks with explicit motion compensation with optical flow.
A faster version without motion compensation, FastDVDnet~\cite{Tassano2020FastDVDnet}, recursively fuses groups of three consecutive frames.
Xue~\etal~\cite{Xue2019TOFlow} train jointly a motion estimation and a video processing network for various specific tasks, including denoising.
More recently, works have begun using attention mechanisms and vision transformers to handle large motions and capture long-range dependencies to aid in video restoration~\cite{Wang2019EDVR,Song2022TempFormer,Liang2024VRT,Liang2022RVRT}.

\subsection{\reviewed{RAW Video Denoising}}

Classical methods can also be easily adapted to work on RAW data by accounting for the CFA structure and the noise model of the sensor~\cite{Chatterjee2011NoiseSuppression,Zhang2009CFA,Patil2016PoissonNR,Zhang2010TwoStage,Akiyama2015Pseudo}.
On video, Zhang~\etal~\cite{Zhang2010SpatialTemporal} apply a spatio-temporal extension of~\cite{Zhang2010TwoStage}, combining patches with the same CFA pattern and thresholding in an adaptive PCA basis.
Kim~\etal~\cite{Kim2015LowLight} perform a motion adaptive temporal filtering on the CFA based on a Kalman structured updating.
Buades and Duran~\cite{Buades2020CFA} extend the method in~\cite{Buades2016Patch} to RAW, packing the CFA into 4 channels, using a variance stabilization transform adapted to an estimated noise curve, and finding a color transformation that maps each frame into a channel-decorrelated space.

Regarding deep learning methods, most are designed to remove Gaussian noise from videos that have already undergone the ISP, and so are unapplicable to the specific structure and noise model of the CFA.
Although many \reviewed{real-noise} denoising models have been proposed in the single image case~\cite{Zhang2018FFDNet,Liu2019Learning,Guo2019CBDNet,Anwar2019RIDNet,Zamir2020CycleISP,Wang2020Practical,Ren2021DeamNet,Gou2022MSANet,Chen2022NAFNet,Zamir2022Restormer}, there are fewer works dealing with RAW video.

The most common RAW video denoising networks are RAW2RAW, meaning they act directly on the CFA and output a denoised version of it.
The majority of them are also supervised, requiring paired data with a clean ground truth, which is hard to obtain for realistic scenarios.
RViDeNet~\cite{Yue2020RViDeNet}
uses deformable convolutions~\cite{Dai2017Deformable} to align the input frames, as well as non-local attention and fusion modules to recover a clean CFA\@.
The authors also provide the CRVD dataset, consisting of stop motion videos of indoor scenes with ground truth and realistic outdoor scenes without one, which they use to train their model.
Paliwal~\etal~\cite{Paliwal2021MaskDnGAN} propose MaskDnGAN, a multi-stage denoiser that registers and fuses small subsequences of overlapping consecutive frames at each stage.
The authors use the unprocessing pipeline by Brooks~\etal~\cite{Brooks2019Unprocessing} to generate synthetic data, and train their model with an adversarial loss, conditioning the discriminator on a soft gradient mask.
Li~\etal~\cite{Li2022Multifrequency} also use a synthetic dataset alongside CRVD to train a method that uses variance stabilization, multi-frame alignment and a multi-frame sequential denoising.
Fu~\etal~\cite{Fu2022LLRVD} collect a dataset that captures low-light scenes with complex motion that are displayed in a monitor, which they use to train a transformer-based network.
Some other supervised networks have been designed for online and unidirectional video denoising~\cite{Maggioni2021EMVD,Ostrowski2022BPEVD,Li2022FloRNN,Qi2022BSVD}.

Sheth~\etal~\cite{Sheth2021UDVD}, inspired by blind-spot networks for still image denoising~\cite{Laine2019BlindSpot,Lehtinen2018Noise2Noise,Batson2019Noise2Self,Krull2019Noise2Void}, propose a CNN that is trained with RAW videos without ground truth.
They use a bias-free architecture and no explicit motion compensation, showing that the network implicitly learns to adapt to local motion.
\reviewed{Relatedly, Wang~\etal~\cite{Wang2023Recurrent} design a recurrent architecture to take advantage of distant temporal information while adjusting to the blind spot requirement.}
Other self-supervised methods for model-blind denoising have been proposed for denoising after demosaicking.
Ehret~\etal~\cite{Ehret2019F2F} construct noisy-noisy pairs by warping neighbouring frames with optical flow, and fine-tune the model by Zhang~\etal~\cite{Zhang2017Beyond}.
Dewil~\etal~\cite{Dewil2021MF2F} follow a similar approach, selecting multiple noisy frames as input.
The benefits of such a paradigm are further studied by Dewil~\etal~\cite{Dewil2022LesserEvil}, who show that training in a self-supervised manner on real noisy data outperforms doing so with supervision on synthetic data.

\subsection{\reviewed{Joint Denoising and Demosaicking}}

\reviewed{
    Much like RAW2RAW, most of the literature on JDD concentrates in the single image case.
    Traditional algorithms can be dated to the work of Hirakawa and Parks~\cite{Hirakawa2006JDD}, who established a theoretical framework for combining both operations.
    Since then, techniques using space-varying filters~\cite{Menon2009JDD}, locally linear embeddings~\cite{Chatterjee2011JDD}, regression tree fields~\cite{Khashabi2014JDD}, least squares luma-chroma demultiplexing~\cite{Jeon2013LumaChroma}, as well as variational formulations of the problem~\cite{Condat2012JDD,Heide2014FlexISP,Klatzer2016SEM,Tan2017ADMM} have been suggested for the task.
    Still, for almost a decade, data driven approaches have been preeminent~\cite{Xing2021EndtoEnd,Guo2021JDD,Chen2021JDD,Zhang2022JDD,Guo2023Joint,Liu2020JDD,Ehret2019JDD,Kokkinos2019Iterative,Gharbi2016DeepJoint,Huang2018Lightweight,Li2024JDD,Chen2018LearningSeeDark}.
}

\reviewed{
    The literature on video JDD, however, has been sparse and mainly focused on very dark scenes.
}
Chen~\etal~\cite{Chen2019SMID} present a dataset of static, low-light sequences with long exposure ground truth, which they use to train a full image processing pipeline \reviewed{with an added temporal consistency loss to avoid flickering}.
Jiang~\etal~\cite{Jiang2019SMOID} use a co-axis optical system with a beam splitter to capture aligned noisy and clean videos with complex motion simultaneously, and use a \reviewed{modified U-Net} to learn a mapping from RAW to sRGB\@.
Wang~\etal~\cite{Wang2021SDSD} also train a \reviewed{CNN-based} pipeline for denoising and illumination enhancement, capturing clean videos with a smooth translational camera motion using a rail system.
More recently, Dewil~\etal~\cite{Dewil2023RVDD} have studied different design aspects of video JDD networks, finding potential in recurrent networks with motion compensation.
\reviewed{More concretely, they obtain best results with a ConvNeXt-based architecture with feature warping and forward frame information.}

\begin{figure*}
	\centering
	\includegraphics[width=\textwidth]{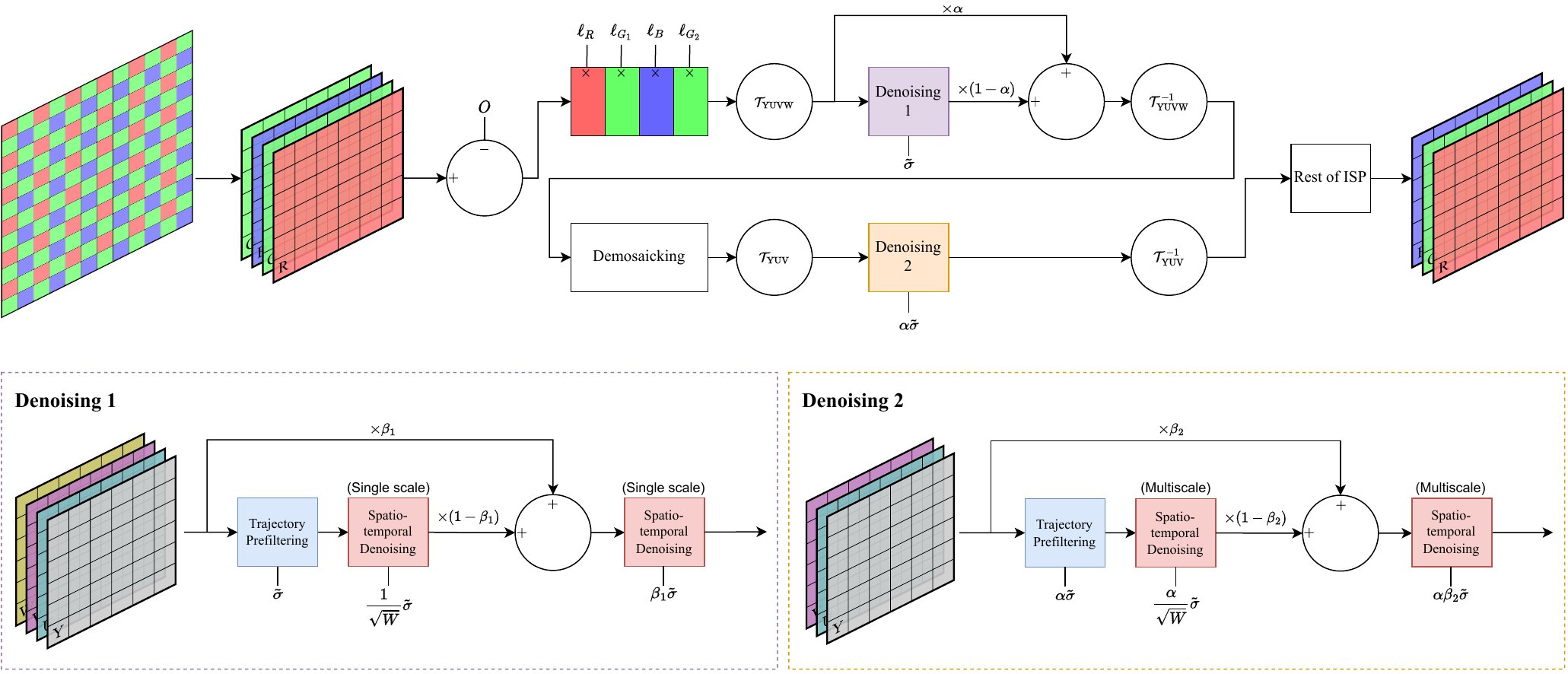}
	\caption{Architecture of the proposed denoising pipeline. We balance two denoising stages, one before and one after demosaicking, returning a portion \(\alpha\) of the noise after the first one. Each stage performs, in a color decorrelated domain, a patch trajectory prefiltering and two iterations of a patch-based spatio-temporal denoising, only requiring an estimation of the sensor noise model.}%
	\label{fig:fullpipeline}
\end{figure*}

\section{Method}

Let \(\mathcal{I} = \{I_{1}, \dotsc, I_{N}\}\) be a video sequence of \(N\) frames in the RAW domain which have not undergone any step of the ISP.
Our proposed pipeline denoises each frame \(I_t \in \mathcal{I}\) sequentially using spatio-temporal information from its surrounding frames \(\{I_{t-T_{b}},\dotsc,I_{t+T_{f}}\}\) with a combination of patch-based strategies with motion compensation.
Instead of having a single denoising block before (or after) demosaicking, we use a modular approach with two denoising blocks at different stages of the ISP.
Namely, we denoise the frames both when they are still mosaicked and after demosaicking, gradually reducing the amount of noise while preserving the details of the scene.
Fig.~\ref{fig:fullpipeline} gives an overall picture of the proposed algorithm architecture.

\subsection{Overall Architecture}

Given the noisy mosaicked \(t\)th image \(I_{t} \in \mathcal{I} \) and a window of its \(W=T_{b}+T_{f}+1\) surrounding frames (including itself), we start by subtracting the black level offset \(O\) added by the sensor and correcting the color through white balance and any other scaling operation in the processing pipeline---like lens shading correction---if present.
We keep negative values so as not to create a bias on the lower tail of the noise distribution.
While not strictly necessary, this early partial application of the ISP is done to help maintain the color constancy condition required by the optical flow and better take advantage of the following color decorrelation transforms used for denoising.

We spatially subsample the frames and pack them into 4-channel images according to their CFA structure.
Following the work in~\cite{Buades2020CFA}, we transform the resulting \(\text{R}\text{G}_{1}\text{B}\text{G}_{2}\) channels into a decorrelated color space YUVW\@.
The transformation \(\mathcal{T}_{\text{YUVW}}\) is reversed after denoising in the RAW domain and before demosaicking.
Analogously, a similar transformation \(\mathcal{T}_{\text{YUV}}\) is used when performing the second denoising stage immediately after demosaicking.

The 4-channel packing in the CFA essentially causes RAW denoising to be performed on an aliased image with a resolution twice lower than the original on each dimension.
Therefore, the denoising process risks oversmoothing image textures and increasing aliasing.
On the other hand, denoising purely after demosaicking has its own drawbacks: the noise is color and spatially correlated, which difficults its removal, and no reliable noise model is available, especially at low signal to noise ratios.
We instead propose a combined two-stage denoising procedure.
After a first denoising stage on the packed CFA data in the YUVW domain (Denoising 1 in Fig.~\ref{fig:fullpipeline}), we return a fraction \(\alpha\) of the noise---defined as the difference between the noisy and the denoised image.
That way, we restore details that may have been oversmoothed, and we facilitate a more precise second denoising stage (Denoising 2 in Fig.~\ref{fig:fullpipeline}) on the now less noisy demosaicked frames.

We use the same structure for both denoising stages.
These are composed of three sub-blocks: an initial temporal prefiltering, a first standard patch-based spatio-temporal denoising, and a partial noise return along with a second iteration of this latter algorithm.
Unlike works like~\cite{Buades2016Patch}, we do not use the result from the first iteration as an oracle for the second one.

The temporal prefiltering sub-block, detailed in Section~\ref{sec:prefiltering}, estimates the trajectories of small overlapping patches and filters their contents, permitting a significant reduction of the noise without producing any texture oversmoothing.
Given \(W\) frames, it approximately decreases the standard deviation of the noise to \(1/\sqrt{W}\) of the original value.
This initial noise reduction facilitates the motion estimation and patch selection in the following sub-blocks.

For the second and third sub-blocks, we carry out two identical applications of a standard PCA-based approach like the one in~\cite{Buades2020CFA}---described in Section~\ref{sec:svddenoising} for completeness.
Between them, we return a fraction of the estimated noise: \(\beta_{1}\) for the first denoising stage, and \(\beta_{2}\) for the second one.

Note that, on the second denoising stage, we estimate the noise standard deviation curve in the demosaicked frames as the one given by the sensor model scaled by \(\alpha\).
As pointed out in~\cite{Jin2020Dilemma}, this is a fair assumption provided that we use stronger filtering thresholds in its sub-blocks.
Apart from that, on this stage we use a multi-scale version of the last two sub-blocks to help remove low-frequency spatially correlated noise.

After the second denoising stage, we proceed with the remaining steps of the ISP.

\subsection{Sensor Noise Model}

The primary sources of noise in a camera sensor are shot noise and read noise.
They can be modeled, respectively, with a Poisson and a Gaussian distribution.
The sensor, moreover, adds a constant positive offset \(O\) to the values it reads in order to avoid the presence of negative values due to noise.
\reviewed{By using the normal approximation of the Poisson distribution}, the noise model of a read pixel \(x_{\text{read}}\) is thus usually described as
\begin{equation}\label{eq:noise_model}
	x_{\text{read}} \sim \mathcal{N}(x_{\text{true}} + O, ax_{\text{true}} + b),
\end{equation}
where \(x_{\text{true}}\) is the true pixel value, and \(a\) and \(b\) depend on the sensor characteristics and on the ISO value used to capture the image~\cite{Foi2008NoiseModeling,Healey1994CCD}.

Given a camera sensor, the noise parameters \(a\) and \(b\) in~\eqref{eq:noise_model} for a particular ISO can be estimated by capturing flat-field frames and bias frames in specialized settings and fitting the observations to a linear model~\cite{Wei2020Physics}.
When dealing with videos in the wild, however, the noise model of the sensor is usually unknown.
Even so, it is possible to estimate the noise distribution on the input set of images, as seen in~\cite{Buades2020CFA}, which adapts the method proposed by Colom~\etal~\cite{Colom2014Nonparametric} and Ponomarenko~\etal~\cite{Ponomarenko2007AVIRIS}.
Consequently, we henceforth assume that the noise model \(\tilde{\sigma}\) is known at the CFA\@.

We do not use any variance stabilization transform.
For videos taken in low light conditions and with high ISO values---and, thus, high noise---it is common to measure intensity values well below the black level offset \(O\).
The estimation of the noise variance given by the model's curve in these regions is usually not as reliable, taking values very close to \(0\).
As noted by Mäkitalo and Foi~\cite{Makitalo2014NoiseMismatch}, applying a typical variance stabilization transformation in such conditions may lead to a loss of detail and a color bias.
In addition, the inverse transform may stretch the dark intensity range, enhancing the residual noise in each block of our pipeline.

Instead, we compute a local variance estimator for each patch neighbourhood while denoising it.
This estimator is computed as the mean of the pixel variances of the reference patch in the neighbourhood.
That is, given a patch \(P\) with pixel positions in \(\{(i_{1}, j_{1}), \dotsc, (i_{r^{2}}, j_{r^{2}})\}\), we estimate the channelwise variance of its neighbourhood \(\mathcal{P}\) as
\begin{equation}\label{eq:noise_patch}
	\bar{\sigma}_{c}^{2} = \frac{1}{r^{2}} \sum_{k=1}^{r^{2}} \ell_{c}^{2} \tilde{\sigma}^{2}\left(\frac{x_{c}(i_{k}, j_{k})}{\ell_{c}}\right),
\end{equation}
where \(x_{c}\) is the intensity of the pixel in channel \(c\), and \(\ell_{c}\) is the coefficient scaling the values in channel \(c\) at the beginning of the imaging pipeline, including white balance and other possibly position-dependent scaling operations.
Note that the application of such position-dependent operations before denoising prevents the use of any variance stabilization transform, which supposes a global noise model.

Since we also use color decorrelation transformations before our denoising blocks, the variance in the transformed domain follows from~\eqref{eq:noise_patch} as
\begin{equation}
	\hat{\sigma}_{c}^{2} = \sum_{k \in C} \mathcal{T}_{c,k}^{2} \bar{\sigma}_{k}^{2},
\end{equation}
where \(\mathcal{T}_{c,k}\) is the coefficient of the transformation matrix \(\mathcal{T}\) for the transformed channel \(c\) and original channel \(k\), and \(C\) is the set of original channels.

\subsection{Temporal Trajectory Prefiltering Block}%
\label{sec:prefiltering}

\begin{figure*}
	\centering
	\includegraphics[width=\textwidth]{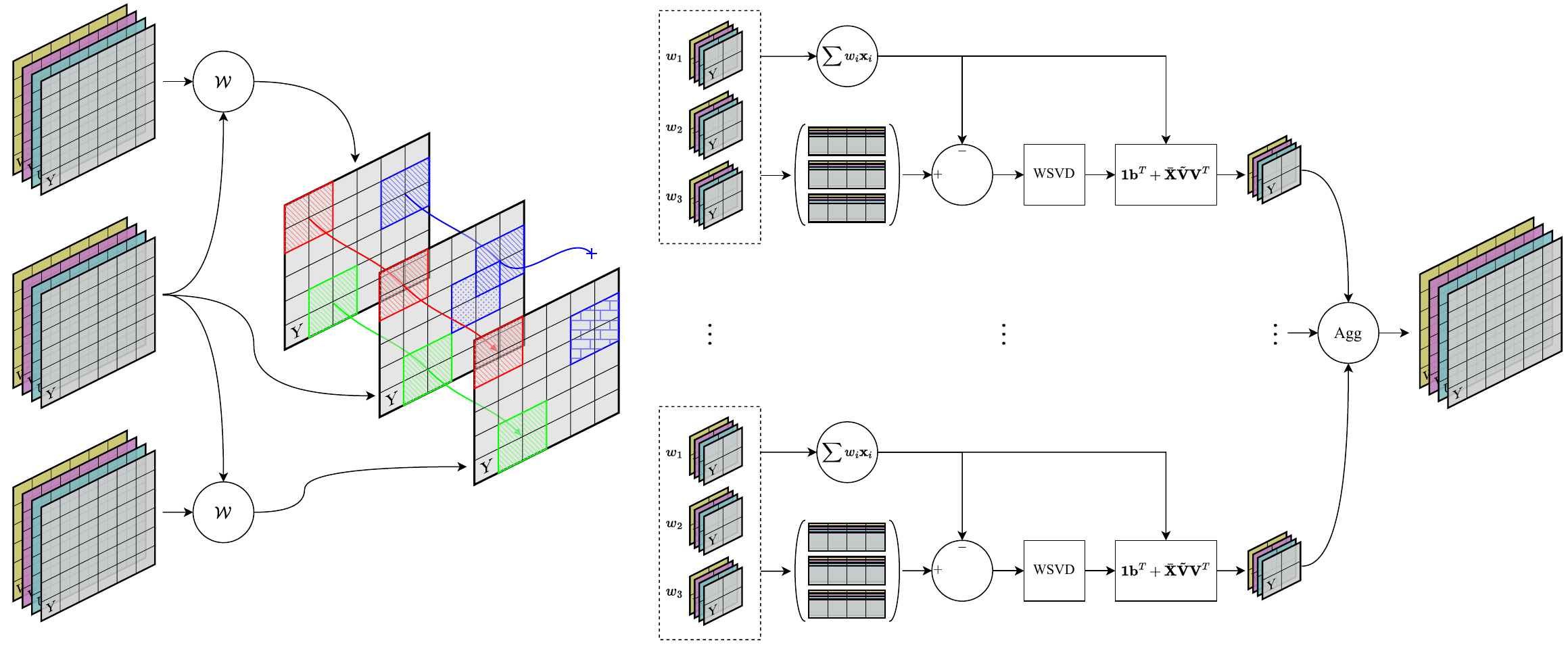}
	\caption{Diagram of the trajectory prefiltering sub-block. Patch trajectories are collected motion-warping adjacent frames and replacing occluded patches with neighbouring ones. The trajectories are filtered by thresholding in a weighted SVD domain, and they are all finally aggregated back into an image.}%
	\label{fig:prefiltering}
\end{figure*}

To exploit temporal redundancy, we establish motion trajectories of patches over a temporal neighbourhood around the reference \(I_{t} \in \mathcal{I}\) to be denoised \reviewed{(Fig.~\ref{fig:prefiltering})}.
We do so by computing the optical flow between \(I_{t}\) and every frame \(I_{i} \in \{I_{t-T_{b}},\dotsc,I_{t+T_{f}}\}\), which is used to spatially align them.
We also compute the inverse flows to check their consistency; if the optical flow and its inverse are not approximately reciprocal at a certain pixel position, we tag it as occluded.

For every \(r \times r\) patch \(P\) in \(I_{t}\), if it has not been occluded in any of the frames, collecting patches in its trajectory yields a set \(\mathcal{P}\) of \(W = T_{b}+T_{f}+1\) patches with almost-equal intensities but different noise realizations.
In this case, simply averaging the patches in \(\mathcal{P}\) would result in a denoised patch with its noise variance reduced by a factor of \(W\).
However, if occluded patches are present in the set, this averaging will produce an incorrectly reconstructed patch.
To avoid this, we replace the occluded patches in \(\mathcal{P}\) with similar non-occluded ones in their spatio-temporal neighbourhood.
That way, this new occlusion-free set \(\hat{\mathcal{P}}\) still contains \(W\) patches that are similar to \(P\).

The patches in \(\hat{\mathcal{P}}\) can still be considerably different.
Averaging them as-is can lead to blurring and loss of detail in regions surrounding fast-moving objects.
To make the prefiltering more robust to optical flow inaccuracies, we instead perform a Weighted Principal Component Analysis (WPCA) on \(\hat{\mathcal{P}}\).

For each \(Q_{i} \in \hat{\mathcal{P}}\), \(i \in \{1,\dotsc,W\}\), we define the weights
\begin{equation}
	w_{i}^{c} = \exp\left( - \frac{\lVert P^{c} - Q_{i}^{c} \rVert^{2}}{h^{2}\hat{\sigma}^{2}_{c}} \right),
\end{equation}
where \(h\) is a parameter, \(\lVert \cdot \rVert\) is the Frobenius norm and the superscript \(c\) denotes the \(c\)th channel of the patches.
Fixed \(c\) and applying the same process to every channel, let \(\mathbf{X}\) be the matrix whose \(i\)th row is comprised of the \(r^{2}\) flattened pixel values of \(Q_{i}^{c}\).
The weighted average
\begin{equation}
	\mathbf{b} = \frac{1}{\sum_{i=1}^{W} w_{i}^{c}} \sum_{i=1}^{W} w_{i}^{c} \mathbf{x}_{i},
\end{equation}
where \(\mathbf{x}_{i}^T\) is the \(i\)th row of \(\mathbf{X}\), is an estimator of the noise-free values of \(P^{c}\) that disfavors unreliable patches that are too different from it.
Despite that, using only \(\mathbf{b}\) to reconstruct it can still result in a blurry patch.
As an alternative, we use \(\mathbf{b}\) to center each row of \(\mathbf{X}\) and consider the eigendecomposition of the weighted covariance matrix
\begin{equation}\label{eq:pca}
	\frac{V_{1}}{V_{1}^{2} - V_{2}} \bar{\mathbf{X}}^{T}\mathbf{W}\bar{\mathbf{X}} = \mathbf{V}\mathbf{\Lambda}\mathbf{V}^{T},
\end{equation}
where \(\mathbf{W} = \operatorname{diag}(w_{1}^{c},\dotsc,w_{W}^{c})\), \(\bar{\mathbf{X}}\) is the centered matrix, and
\begin{equation}
	V_{1} = \sum_{i=1}^{W} w_{i}^{c}, \qquad V_{2} = \sum_{i=1}^{W} (w_{i}^{c})^{2}.
\end{equation}

Our objective is to reconstruct a filtered version of \(P^{c}\) by adding detail---but not noise---to \(\mathbf{b}\).
We do so by adding to it an information-maximizing low-rank reconstruction of \(\bar{\mathbf{X}}\) obtained by keeping only the first few eigenvectors in \(\mathbf{V}\),
\begin{equation}\label{eq:reconstruction}
	\hat{\mathbf{X}} = \mathbf{1}_{W \times 1}\mathbf{b}^{T} + \bar{\mathbf{X}}\tilde{\mathbf{V}}\mathbf{V}^{T},
\end{equation}
thresholding them according to the magnitude of their respective eigenvalue, and retrieving the row corresponding to \(P^{c}\).
Particularly, we set \(\tilde{\mathbf{V}} = \mathbf{V}\tilde{\mathbf{D}}\), where \(\tilde{\mathbf{D}}\) is diagonal with
\begin{equation}\label{eq:thresholdmatrix}
	\tilde{D}_{ii} =
	\begin{cases}
		1 & \text{if } \lambda_{i} \geqslant \delta_{i}, \\
		0 & \text{otherwise},
	\end{cases}
\end{equation}
and \(\delta_{i}\) is a parameter that fixes a cutoff value for the variance explained by the \(i\)th component of the PCA\@.
To compute \(\mathbf{V}\) and \(\mathbf{\Lambda}\) we perform the singular value decomposition of \(\mathbf{W}^{\frac{1}{2}}\bar{\mathbf{X}}\) and use the relation \(\mathbf{\Lambda} = \frac{V_{1}}{V_{1}^{2} - V_{2}}\mathbf{S}^{2}\) between the eigenvalues and the singular values to set the threshold in~\eqref{eq:thresholdmatrix}.
Note that in an ideal case without occlusions and with perfect optical flow, all the signal information needed to denoise the patch would be contained in \(\mathbf{b}\), so the last term in~\eqref{eq:reconstruction} would be filtered out completely by the threshold~\eqref{eq:thresholdmatrix}, leading to a simple averaging of the trajectory.
In practice, that term will add important signal information with minimal noise.

\subsection{Spatio-Temporal Denoising Block}%
\label{sec:svddenoising}

The foundation of the spatio-temporal denoising sub-block used in the proposed pipeline is the PCA-based denoising algorithm proposed in~\cite{Buades2020CFA}.
Like in the temporal prefiltering sub-block, we warp the frames in \(\{I_{t-T_{b}},\dotsc,I_{t+T_{f}}\}\) into \(I_{t}\) and compute an occlusion mask for each pixel using the optical flow between the frames.
Then, we use a 3D volumetric approach to search for blocks of similar patches.

For each overlapping patch \(P\) in \(I_{t}\), we again consider its temporal extension \(\mathcal{P}\).
Instead of working directly on the patches in \(\mathcal{P}\), now we search its vicinity for the \(K\) extended patches \(\{\mathcal{Q}_{1},\dotsc,\mathcal{Q}_{K}\}\) most similar to it using the mean patchwise Euclidean distance.
Patches with occluded pixels, either in \(\mathcal{P}\) or in any \(\mathcal{Q}_{i}\), are discarded and not taken into account for the similarity computation.
Working on already prefiltered frames with a fraction of the original sensor noise permits to mitigate its influence on the patch selection and to improve the accuracy of the search.
The selected 3D volumes are then sliced in order into their constituent 2D patches until a total of \(M\) patches is reached, where \(M \geqslant r^{2}\).
The PCA of the block formed by these patches is computed channelwise, and their denoised counterparts are obtained by canceling the principal coefficients whose respective eigenvalues are lower than \(\tau_{c}^{2}\hat{\sigma}_{c}^{2}\) for a fixed \(\tau_{c}\).

After denoising the block of patches, we not only retrieve the denoised version of patch \(P\), but also the rest of the patches that lie on the same frame \(I_{t}\).
Because of the overlap between the different blocks that can be built by sliding the reference patch \(P\) along the image, the collection of all the denoised patches form an overcomplete representation of it.
To obtain the final denoised frame, we aggregate the denoised patches by a weighted averaging at the positions where pixels are shared.
Similarly to BM3D~\cite{Dabov2007BM3D}, we use aggregation weights inversely proportional to the number of retained PCA coefficients, and use a Kaiser window to reduce ringing artifacts.

\subsection{Implementation Details}

\textbf{Color decorrelation transformations.} Standard orthogonal color transformations used for denoising, under the assumption that the noise has the same amplitude in each channel, use the same proportion of each color to compute the luma.
In our case, because we apply white balance before denoising and do not use a VST, this assumption does not hold.
We thus choose to sacrifice orthogonality in favor of a luma representation that gives more significance to green.
At the same time, we keep the zero-sum row property for the chromatic components so that gray pixels are represented with null chroma values in the new space.
The transformation matrices we use for \(\text{R}\text{G}_{1}\text{B}\text{G}_{2} \rightarrow \text{Y}\text{U}\text{V}\text{W}\) and \(\text{R}\text{G}\text{B} \rightarrow \text{Y}\text{U}\text{V}\) are then, respectively,
\begin{equation}
	\mathcal{T}_{\text{YUVW}} = \begin{pmatrix}
		0.3162  & 0.65   & 0.2345 & 0.65    \\
		-0.5    & 0.5    & -0.5   & 0.5     \\
		0.65    & 0.2784 & -0.65  & -0.2784 \\
		-0.2784 & 0.65   & 0.2784 & -0.65
	\end{pmatrix}
\end{equation}
and
\begin{equation}
	\mathcal{T}_{\text{YUV}} = \begin{pmatrix}
		0.299  & 0.587  & 0.114  \\
		-0.147 & -0.289 & 0.436  \\
		0.615  & -0.515 & -0.100
	\end{pmatrix}.
\end{equation}

\medskip

\textbf{Optical flow.} We use the TV-\(L^{1}\) optical flow algorithm~\cite{Zach2007TVL1} on the Y channel to compute the motion between frames.
We use a data fidelity term \(\lambda=0.7\) and an inverse flow reciprocity distance of \(0.25\) pixels as the limit before a pixel is invalidated.

\medskip

\textbf{Demosaicking.}
\reviewed{We demosaic the CFA data with the simple algorithm from Duran and Buades presented in~\cite[Section II]{Duran2014Demosaicking}.}
\reviewed{This method reconstructs the color image by a weighted average of four directional interpolations.  
Each local interpolation is obtained by applying a Hamilton-Adams~\cite{Hamilton1997HA} stencil in a particular direction. 
The combination depends on the chromatic variation of the interpolated image in the corresponding direction.  
}

\medskip

\textbf{Balance between denoisers.} The optimal proportion \(\alpha\) of returned noise before demosaicking should depend on the conditions of the captured scene.
Low-light scenes taken with a higher ISO benefit from a lower \(\alpha\)---that is, a stronger denoising before demosaicking---, as the noise model in darker regions is more reliable in the RAW domain and the correlated noise after demosaicking can be hard to remove due to the lower signal-to-noise ratio.
On the contrary, brighter, low-ISO videos can afford to give more influence to the second denoising step to improve texture recovery.
We study the effect of \(\alpha\) in more detail in Section~\ref{sec:ablation}.

Unlike \(\alpha\), the parameters \(\beta_{1}\) and \(\beta_{2}\) of returned noise between iterations in the same image domain are not influenced by external factors.
We set them to a constant \(\beta_{1}=\beta_{2}=0.3\).

\medskip

\textbf{Denoising parameters.} In all stages of the pipeline we use square patches of size \(7 \times 7\).
Both temporal prefiltering blocks are performed on a single scale, with a weighting parameter \(h=1.25r=8.75\).
We choose \(\delta_{i} = (1.25\hat{s}_{i})^{2}V_{1}/(V_{1}^{2}-V_{2})\), where \(\hat{s}_i\) is the expectation of the \(i\)th singular value of the noise component of the weighted trajectory matrix (with noise variance \(\hat{\sigma}_{c}^{2}\)), which can be assumed to follow a Marchenko-Pastur probability distribution.
This expectation is computed as in eqs.\ (120) and (121) in~\cite{Epps2019SVD}, using the spatial correlation factor the authors propose to compensate for the effects of the weights on the data.

For the spatio-temporal denoising blocks, we fix \(K=66\) neighbours.
In the RAW domain, we use a single scale with \(\tau_{Y}=1.9\) for the luminosity and \(\tau_{c}=2.2\) for the chromaticity components.
After demosaicking, the number of scales is increased to three.
We set a slighly higher threshold for the chromaticity components on the coarsest one (\(0.8\) vs.\ \(0.6\) for Y) to help remove low frequency chromatic noise.
The other scales use the same threshold value for each channel, which we fix at \(1.0\) for the middle one and \(3.0\) for the finest.
Note that this last value is close to \(1.5\) times the one used before demosaicking, which coincides with the optimal multiplicative factor described by~\cite{Jin2020Dilemma} for denoising demosaicked data.

\section{Experimental results and discussion}

In this section, we study the performance of the proposed method.
First, we perform an ablation study and discuss the impact of the different components that it comprises.
We look into the optimal balance between the two denoising stages, and we analyze the effect of the trajectory prefiltering sub-blocks, as well as that of the second iteration of the spatio-temporal denoising sub-blocks \reviewed{and the choice of demosaicking algorithm}.
Finally, we make qualitative and quantitative comparisons with state-of-the-art methods for RAW video denoising on data with realistic sensor noise.

\subsection{Datasets, Visualization, and Metrics}

We use three different datasets for our experiments.
For qualitative comparisons, we use the outdoor video sequences in the CRVD dataset~\cite{Yue2020RViDeNet}, for which ground truth is not available. 
\reviewed{These are real noisy video RAW sequences.}
We choose to dismiss the indoor sequences from that dataset because most of them are used as training data by deep learning methods.

For numerical comparisons, we use the synthetic test set used by Paliwal~\etal~\cite{Paliwal2021MaskDnGAN}.
The dataset consists of 12 dynamic scenes whose frames have been unprocessed using the chain in~\cite{Brooks2019Unprocessing}.
A noise realization is generated for each frame with the noise model of the sensor used in CRVD, for ISOs ranging from 1600 to 25600.
\reviewed{While the unprocessing chain~\cite{Brooks2019Unprocessing} is widely used, we must emphasize that these are not real RAW video sequences.}

Lastly, we add CRVD-like noise to sequences in the ARRI dataset~\cite{Andriani2013ARRI} for \reviewed{four} more ISO values, interpolating the noise model's parameters.
\reviewed{These are very high signal-to-noise ratio real RAW sequences which can be considered to be noise-free.} 
We use them for additional qualitative and quantitative comparisons with the state of the art \reviewed{on real data}.

Throughout the experiments, we use the processing pipeline illustrated in Fig.~\ref{fig:fullpipeline} to visualize the images.
The modules of the ISP are chosen depending on the dataset.
For CRVD and the synthetic test set from Paliwal~\etal~\cite{Paliwal2021MaskDnGAN}, we use CRVD's own white balance coefficients and camera correction matrix, provided by the authors, along with a gamma correction with \(\gamma=2.2\).
For ARRI, we use the processing Matlab script that comes with the dataset, which includes a white balance and other, more complex corrections.

To guarantee that our quantitative assessments reflect the quality of a video as it is visualized, all measurements are taken on the sRGB images after the full processing pipeline.
To obtain all of them, we use the FFmpeg software to build video files in YUV 4:2:0 format at 24 FPS, which we feed to Netflix's official implementation of VMAF~\cite{Li2016VMAF}.
We extract the PSNR, SSIM~\cite{Wang2004SSIM}, MS-SSIM~\cite{Wang2003MSSSIM}, and VMAF~\cite{Li2016VMAF} scores from the output of the software.

\subsection{Ablation Study}%
\label{sec:ablation}

\begin{figure*}
    \centering
    \begin{minipage}{0.1644\linewidth}%
        \subfloat[Scene 7 (ISO 1600)]{%
            \includegraphics[width=\linewidth]{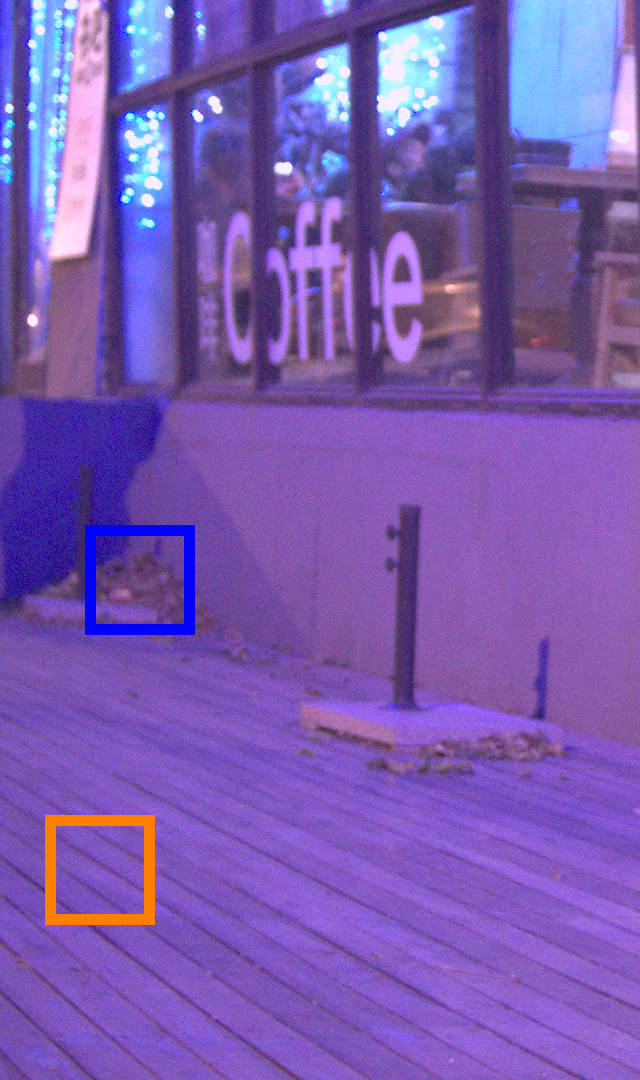}%
        }%
    \end{minipage}\hfill%
    \begin{minipage}{0.8256\linewidth}%
        \includegraphics[width=0.158\linewidth,cframe=blue 1pt]{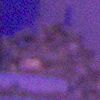}%
        \hfill%
        \includegraphics[width=0.158\linewidth,cframe=blue 1pt]{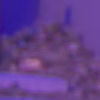}%
        \hfill%
        \includegraphics[width=0.158\linewidth,cframe=blue 1pt]{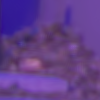}%
        \hfill%
        \includegraphics[width=0.158\linewidth,cframe=blue 1pt]{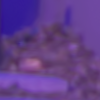}%
        \hfill%
        \includegraphics[width=0.158\linewidth,cframe=blue 1pt]{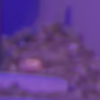}%
        \hfill%
        \includegraphics[width=0.158\linewidth,cframe=blue 1pt]{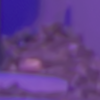}%
        \\[-0.5em]%
        \subfloat[Noisy]{%
            \includegraphics[width=0.158\linewidth,cframe=orange 1pt]{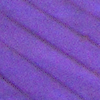}%
        }\hfill%
        \subfloat[\(\alpha=0\)]{%
            \includegraphics[width=0.158\linewidth,cframe=orange 1pt]{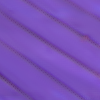}%
        }\hfill%
        \subfloat[\(\alpha=0.3\)]{%
            \includegraphics[width=0.158\linewidth,cframe=orange 1pt]{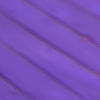}%
        }\hfill%
        \subfloat[\(\alpha=0.5\)]{%
            \includegraphics[width=0.158\linewidth,cframe=orange 1pt]{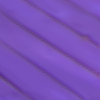}%
        }\hfill%
        \subfloat[\(\alpha=0.7\)]{%
            \includegraphics[width=0.158\linewidth,cframe=orange 1pt]{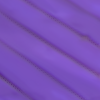}%
        }\hfill%
        \subfloat[\(\alpha=1\)]{%
            \includegraphics[width=0.158\linewidth,cframe=orange 1pt]{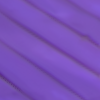}%
        }
    \end{minipage}%
    \par\bigskip%
    \begin{minipage}{0.1644\linewidth}%
        \subfloat[Scene 7 (ISO 25600)]{%
            \includegraphics[width=\linewidth]{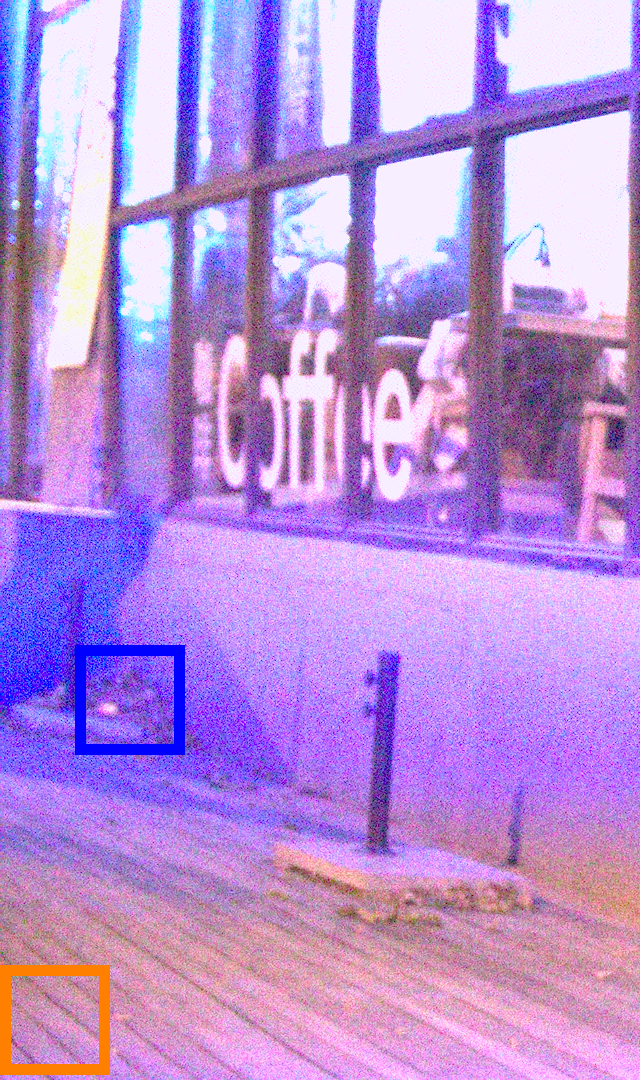}%
        }%
    \end{minipage}\hfill%
    \begin{minipage}{0.8256\linewidth}%
        \includegraphics[width=0.158\linewidth,cframe=blue 1pt]{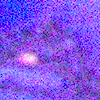}%
        \hfill%
        \includegraphics[width=0.158\linewidth,cframe=blue 1pt]{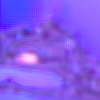}%
        \hfill%
        \includegraphics[width=0.158\linewidth,cframe=blue 1pt]{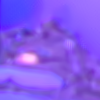}%
        \hfill%
        \includegraphics[width=0.158\linewidth,cframe=blue 1pt]{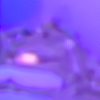}%
        \hfill%
        \includegraphics[width=0.158\linewidth,cframe=blue 1pt]{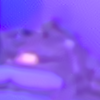}%
        \hfill%
        \includegraphics[width=0.158\linewidth,cframe=blue 1pt]{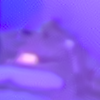}%
        \\[-0.5em]%
        \subfloat[Noisy]{%
            \includegraphics[width=0.158\linewidth,cframe=orange 1pt]{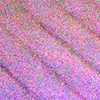}%
        }\hfill%
        \subfloat[\(\alpha=0\)]{%
            \includegraphics[width=0.158\linewidth,cframe=orange 1pt]{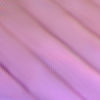}%
        }\hfill%
        \subfloat[\(\alpha=0.3\)]{%
            \includegraphics[width=0.158\linewidth,cframe=orange 1pt]{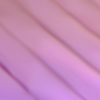}%
        }\hfill%
        \subfloat[\(\alpha=0.5\)]{%
            \includegraphics[width=0.158\linewidth,cframe=orange 1pt]{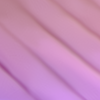}%
        }\hfill%
        \subfloat[\(\alpha=0.7\)]{%
            \includegraphics[width=0.158\linewidth,cframe=orange 1pt]{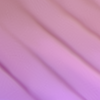}%
        }\hfill%
        \subfloat[\(\alpha=1\)]{%
            \includegraphics[width=0.158\linewidth,cframe=orange 1pt]{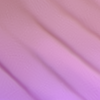}%
        }
    \end{minipage}%
    \caption{Visual quality comparison with varying values of \(\alpha\). No prefiltering nor iteration sub-blocks are used. Denoising purely before demosaicking (\(\alpha=0\)) leaves checkerboard artifacts, while doing so purely after (\(\alpha=1\)) leaves spatially correlated residual noise. \reviewed{A combination of both prevents both problems.} Higher ISOs have a lower optimal \(\alpha\) value. \reviewed{Scene from the CRVD outdoor dataset.}}%
    \label{fig:ablationalpha}
\end{figure*}

\begin{table}
    \centering
    \caption{Effect of varying \(\alpha\) on the proposed method. No prefiltering nor iteration sub-blocks used. Measures taken on MaskdnGAN's test set. }%
    \label{tab:ablationalpha}
    \begin{tabular}{clcccc}
        \toprule
		ISO & \(\alpha\) & PSNR & SSIM & MS-SSIM & VMAF \\
		\midrule
		\multirow{5}{*}{1600}  & 0   & 39.33             & 0.9970             & 0.9941             & 98.63             \\
                               & 0.3 & 40.35             & \textbf{0.9972}    & \underline{0.9950} & \underline{98.79} \\
		                       & 0.5 & \textbf{40.69}    & \textbf{0.9972}    & \textbf{0.9951}    & \textbf{98.84}    \\
                               & 0.7 & \underline{40.62} & 0.9970             & 0.9948             & 98.78             \\
		                       & 1   & 40.11             & 0.9961             & 0.9936             & 98.51             \\
		\midrule
		\multirow{5}{*}{3200}  & 0   & 37.18             & 0.9950             & 0.9904             & 97.75             \\
                               & 0.3 & 38.34             & \textbf{0.9954}    & \textbf{0.9918}    & \underline{98.03} \\
                               & 0.5 & \textbf{38.65}    & \underline{0.9952} & \textbf{0.9918}    & \textbf{98.06}    \\
                               & 0.7 & \underline{38.52} & 0.9946             & 0.9910             & 97.87             \\
		                       & 1   & 37.87             & 0.9925             & 0.9883             & 97.28             \\
		\midrule
        \multirow{5}{*}{6400}  & 0   & 35.18             & \underline{0.9916} & 0.9844             & 95.48             \\
                               & 0.3 & \underline{36.39} & \textbf{0.9921}    & \textbf{0.9866}    & \textbf{96.51}    \\
                               & 0.5 & \textbf{36.61}    & 0.9915             & \underline{0.9861} & \underline{96.50} \\
		                       & 0.7 & 36.35             & 0.9898             & 0.9839             & 95.92             \\
		                       & 1   & 35.49             & 0.9848             & 0.9778             & 93.46             \\
		\midrule
        \multirow{5}{*}{12800} & 0   & 33.23             & \underline{0.9853} & 0.9744             & 88.15             \\
                               & 0.3 & \underline{34.37} & \textbf{0.9857}    & \textbf{0.9772}    & \textbf{90.47}    \\
                               & 0.5 & \textbf{34.40}    & 0.9837             & \underline{0.9749} & \underline{90.19} \\
		                       & 0.7 & 33.96             & 0.9793             & 0.9695             & 87.95             \\
		                       & 1   & 32.85             & 0.9679             & 0.9564             & 81.67             \\
		\midrule
        \multirow{5}{*}{25600} & 0   & 31.39             & \textbf{0.9732}    & \underline{0.9579} & 77.32             \\
                               & 0.3 & \textbf{32.30}    & \underline{0.9725} & \textbf{0.9599}    & \textbf{79.83}    \\
                               & 0.5 & \underline{32.09} & 0.9669             & 0.9533             & \underline{78.32} \\
		                       & 0.7 & 31.47             & 0.9569             & 0.9420             & 74.07             \\
		                       & 1   & 30.27             & 0.9339             & 0.9183             & 64.50             \\
		\bottomrule
	\end{tabular}
\end{table}

Unless otherwise stated, to perform the experiments in this section, we fix \(T_{b}=T_{f}=1\) as the number of backward and forward frames to feed to each denoising sub-block.
We first study the effect of balancing denoising pre- and post-demosaicking.
We disable the prefiltering and spatio-temporal iteration sub-blocks, and vary the parameter \(\alpha\) from \(0\) (equivalent to fully denoising in the RAW domain) to \(1\) (fully denoising after demosaicking).
Fig.~\ref{fig:ablationalpha} shows the results for a scene from the CRVD outdoor dataset for both a low (1600) and a high (25600) ISO value.
For \(\alpha=0\), both images present some checkerboard artifacts due to effectively being denoised at a lower resolution.
A slight increase in \(\alpha\) already removes these artifacts, significantly improving image quality.
On the other hand, with \(\alpha=1\), some spatially correlated residual noise still remains, which is especially noticeable at higher ISOs and can be reduced by decreasing the value.
Table~\ref{tab:ablationalpha} confirms our observations numerically on the synthetic test set.
The lower noise levels reach a higher PSNR with \(\alpha=0.5\) (\(\alpha=0.7\) being a close second for ISO 1600 and 3200).
This optimal value decreases gradually until reaching \(\alpha=0.3\) for ISO 25600.
\reviewed{A difference in PSNR between \SI{0.5}{\dB} and \SI{2}{\dB} is observed across all ISOs when comparing the optimal balance with the application of a single denoising stage, being especially noticeable ($>\SI{1}{\dB}$) starting from ISO 6400.}
The other metrics show a similar decreasing trend, beginning with an optimal \(\alpha=0.5\) for ISO 1600 and predominantly preferring \(\alpha=0.3\) from ISO 3200 onward.
For the rest of the experiments, we fix \(\alpha\) depending on the ISO, choosing the value that maximizes PSNR in Table~\ref{tab:ablationalpha}.

\begin{table}
    \centering
    \caption{Effect of the chosen demosaicking algorithm. No prefiltering nor iteration sub-blocks used. Measures taken on MaskdnGAN's test set. }%
    \label{tab:ablationdemo}
    \begin{tabular}{clcccc}
        \toprule
		ISO & Demosaicking & PSNR & SSIM & MS-SSIM & VMAF \\
		\midrule
        \multirow{2}{*}{1600}  & Original~\cite{Duran2014Demosaicking} & 40.69 & 0.9972 & 0.9951 & 98.84 \\
                               & MLRI~\cite{Kiku2014MLRI}        & 40.21 & 0.9972 & 0.9950 & 98.83 \\
		\midrule
        \multirow{2}{*}{3200}  & Original~\cite{Duran2014Demosaicking} & 38.65 & 0.9952 & 0.9918 & 98.06 \\
                               & MLRI~\cite{Kiku2014MLRI}        & 38.44 & 0.9953 & 0.9919 & 98.07 \\
		\midrule
        \multirow{2}{*}{6400}  & Original~\cite{Duran2014Demosaicking} & 36.61 & 0.9915 & 0.9861 & 96.50 \\
                               & MLRI~\cite{Kiku2014MLRI}        & 36.55 & 0.9917 & 0.9864 & 96.51 \\
		\midrule
        \multirow{2}{*}{12800} & Original~\cite{Duran2014Demosaicking} & 34.40 & 0.9837 & 0.9749 & 90.19 \\
                               & MLRI~\cite{Kiku2014MLRI}        & 34.46 & 0.9842 & 0.9758 & 90.39 \\
		\midrule
        \multirow{2}{*}{25600} & Original~\cite{Duran2014Demosaicking} & 32.30 & 0.9725 & 0.9599 & 79.83 \\
                               & MLRI~\cite{Kiku2014MLRI}        & 32.30 & 0.9730 & 0.9607 & 79.83 \\
		\bottomrule
	\end{tabular}
\end{table}

\begin{figure*}
    \centering
    \begin{minipage}{0.19\linewidth}%
        \subfloat[Scene 4 (ISO 3200)]{%
            \includegraphics[width=\linewidth]{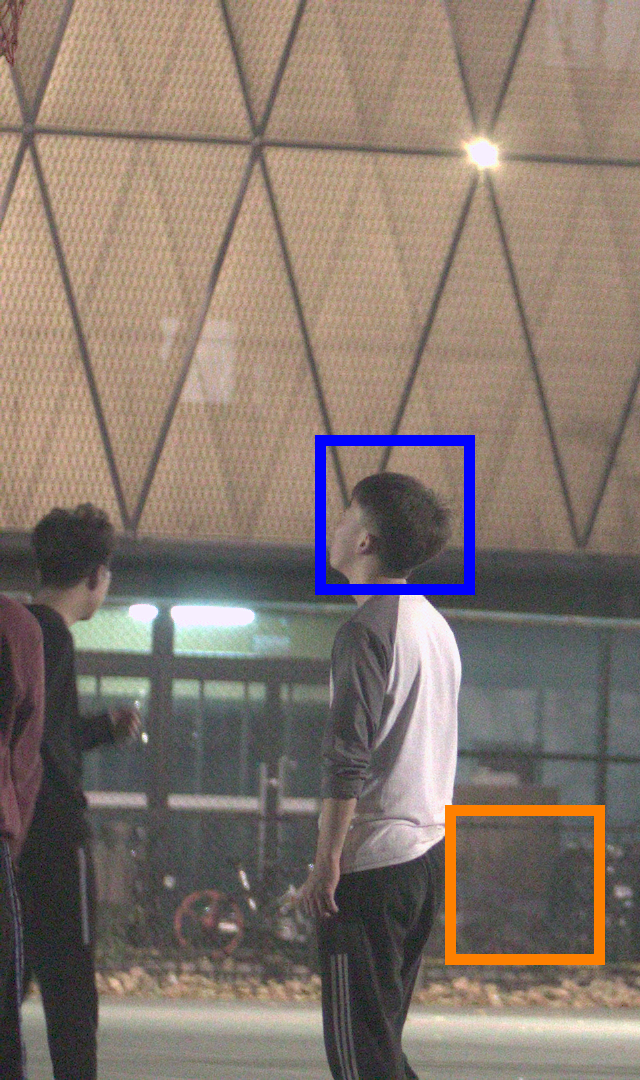}%
        }%
    \end{minipage}\hfill%
    \begin{minipage}{0.8\linewidth}%
        \includegraphics[width=0.19\linewidth,cframe=blue 1pt]{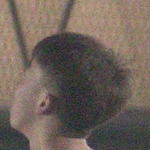}%
        \hfill%
        \includegraphics[width=0.19\linewidth,cframe=blue 1pt]{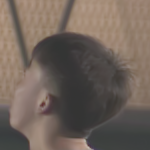}%
        \hfill%
        \includegraphics[width=0.19\linewidth,cframe=blue 1pt]{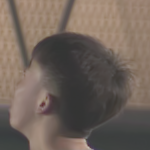}%
        \hfill%
        \includegraphics[width=0.19\linewidth,cframe=blue 1pt]{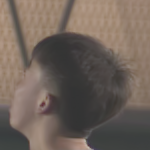}%
        \hfill%
        \includegraphics[width=0.19\linewidth,cframe=blue 1pt]{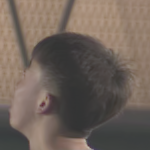}%
        \\[-0.5em]%
        \subfloat[Noisy]{%
            \includegraphics[width=0.19\linewidth,cframe=orange 1pt]{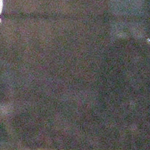}%
        }\hfill%
        \subfloat[No prefiltering]{%
            \includegraphics[width=0.19\linewidth,cframe=orange 1pt]{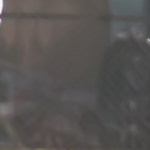}%
        }\hfill%
        \subfloat[Before demo]{%
            \includegraphics[width=0.19\linewidth,cframe=orange 1pt]{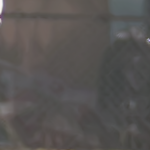}%
        }\hfill%
        \subfloat[After demo]{%
            \includegraphics[width=0.19\linewidth,cframe=orange 1pt]{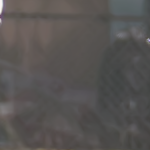}%
        }\hfill%
        \subfloat[Before \& after]{%
            \includegraphics[width=0.19\linewidth,cframe=orange 1pt]{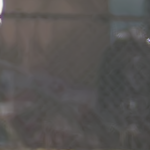}%
        }%
    \end{minipage}%
    \caption{Visual quality comparison with different combinations of prefiltering sub-blocks before and after demosaicking. Spatio-temporal denoising sub-blocks are used before and after demosaicking with \(\alpha=0.5\). No iteration sub-blocks are used. \reviewed{Temporally prefiltering before and after demosaicking allows a better reconstruction of finer details. Scene from the CRVD outdoor dataset.}}%
    \label{fig:ablationprefiltering}
\end{figure*}

\begin{table}
    \centering
    \caption{Effect of the temporal prefiltering sub-blocks. No iteration sub-blocks used. Measures taken on MaskdnGAN's test set.}%
    \label{tab:ablationprefiltering}
	\begin{tabular}{ccccccc}
		\toprule
		\multirow{2}{*}[-0.5\dimexpr\aboverulesep+\belowrulesep+\cmidrulewidth]{ISO} & \multicolumn{2}{c}{Prefiltering} & \multirow{2}{*}[-0.5\dimexpr\aboverulesep+\belowrulesep+\cmidrulewidth]{PSNR} & \multirow{2}{*}[-0.5\dimexpr\aboverulesep+\belowrulesep+\cmidrulewidth]{SSIM} & \multirow{2}{*}[-0.5\dimexpr\aboverulesep+\belowrulesep+\cmidrulewidth]{MS-SSIM} & \multirow{2}{*}[-0.5\dimexpr\aboverulesep+\belowrulesep+\cmidrulewidth]{VMAF}   \\
		\cmidrule(lr){2-3}
		& RAW & Demo & & & & \\
		\midrule
        \multirow{4}{*}{1600} & \xmark & \xmark & 40.69 & 0.9972 & 0.9951 & 98.84 \\
        & \xmark & \cmark & \textbf{40.72} & 0.9974 & 0.9954 & \textbf{98.91} \\
        & \cmark & \xmark & 40.67 & 0.9974 & 0.9952 & 98.86 \\
        & \cmark & \cmark & 40.64 & \textbf{0.9975} & \textbf{0.9955} & 98.90 \\
        \midrule
        \multirow{4}{*}{3200} & \xmark & \xmark & 38.65 & 0.9952 & 0.9918 & 98.06 \\
        & \xmark & \cmark & \textbf{38.84} & 0.9958 & 0.9927 & 98.25 \\
        & \cmark & \xmark & 38.68 & 0.9956 & 0.9922 & 98.16 \\
        & \cmark & \cmark & 38.81 & \textbf{0.9961} & \textbf{0.9930} & \textbf{98.31} \\
        \midrule
        \multirow{4}{*}{6400} & \xmark & \xmark & 36.61 & 0.9915 & 0.9861 & 96.50 \\
        & \xmark & \cmark & 36.93 & 0.9927 & 0.9879 & 96.97 \\
        & \cmark & \xmark & 36.72 & 0.9925 & 0.9871 & 96.82 \\
        & \cmark & \cmark & \textbf{37.01} & \textbf{0.9935} & \textbf{0.9888} & \textbf{97.22} \\
        \midrule
        \multirow{4}{*}{12800} & \xmark & \xmark & 34.40 & 0.9837 & 0.9749 & 90.19 \\
        & \xmark & \cmark & 34.81 & 0.9862 & 0.9784 & 91.78 \\
        & \cmark & \xmark & 34.65 & 0.9857 & 0.9774 & 91.74 \\
        & \cmark & \cmark & \textbf{35.06} & \textbf{0.9879} & \textbf{0.9807} & \textbf{93.27} \\
        \bottomrule
	\end{tabular}
\end{table}

\reviewed{
    To discard that the results are specific to the chosen demosaicking algorithm, we switch it to the Minimized-Laplacian Residual Interpolation (MLRI) method~\cite{Kiku2014MLRI} and reapply the two-stage denoising process.
    The numbers in Table~\ref{tab:ablationdemo} corroborate that there is no significant difference between the two methods, except for low ISO settings, where the original choice of algorithm yields slightly higher PSNR values.
    Differences in SSIM, MS-SSIM and VMAF are marginal and do not justify a change of algorithm.
}

In Fig.~\ref{fig:ablationprefiltering} we re-enable the temporal prefiltering blocks one by one to show their effect on denoising quality.
It can be seen that prefiltering allows a more accurate texture reconstruction, especially on regions with fine details (e.g.~hair or a fence) that otherwise are usually oversmoothed.
Also, although already noticeable at lower ISOs, the impact of temporal prefiltering makes itself more evident as ISO increases, as shown in Table~\ref{tab:ablationprefiltering}.
It is important to note that, if an object on a sequence is difficult to track (i.e.\ it moves fast), an erroneous optical flow estimation may cause reconstruction inaccuracies.
Besides strong flow reciprocity restrictions, our use of a weighted PCA instead of a simple weighted average along the trajectories mitigates this problem.

\begin{figure*}
    \centering
    \begin{minipage}{0.19\linewidth}%
        \subfloat[Scene 5 (ISO 12800)]{%
            \includegraphics[width=\linewidth]{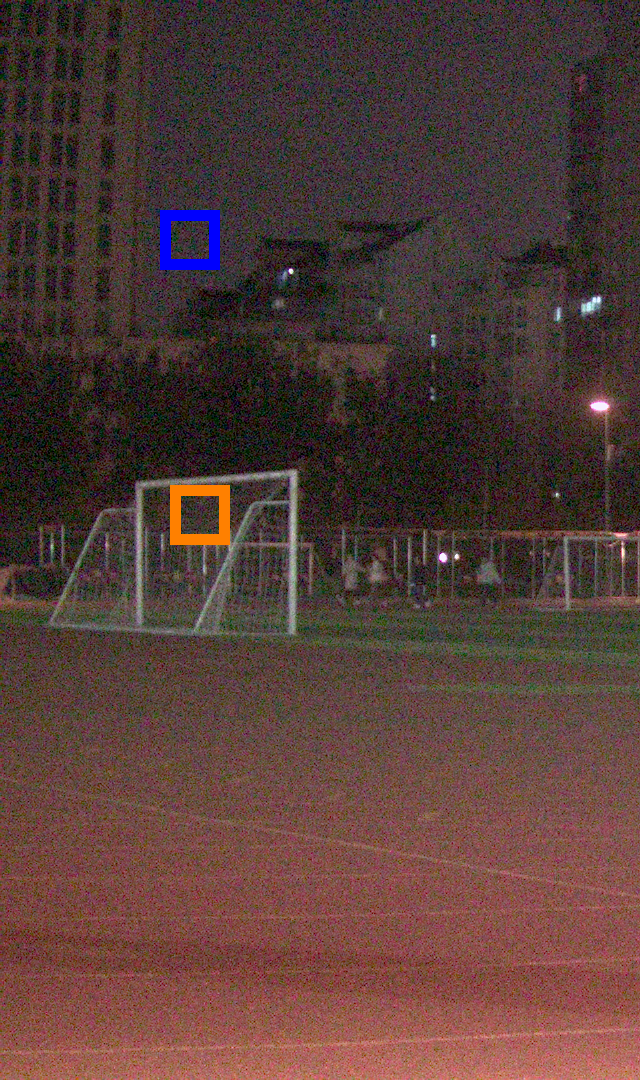}%
        }%
    \end{minipage}\hfill%
    \begin{minipage}{0.8\linewidth}%
        \includegraphics[width=0.19\linewidth,cframe=blue 1pt]{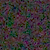}%
        \hfill%
        \includegraphics[width=0.19\linewidth,cframe=blue 1pt]{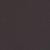}%
        \hfill%
        \includegraphics[width=0.19\linewidth,cframe=blue 1pt]{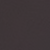}%
        \hfill%
        \includegraphics[width=0.19\linewidth,cframe=blue 1pt]{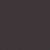}%
        \hfill%
        \includegraphics[width=0.19\linewidth,cframe=blue 1pt]{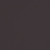}%
        \\[-0.5em]%
        \subfloat[Noisy]{%
            \includegraphics[width=0.19\linewidth,cframe=orange 1pt]{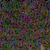}%
        }\hfill%
        \subfloat[No iteration]{%
            \includegraphics[width=0.19\linewidth,cframe=orange 1pt]{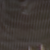}%
        }\hfill%
        \subfloat[Before demo]{%
            \includegraphics[width=0.19\linewidth,cframe=orange 1pt]{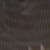}%
        }\hfill%
        \subfloat[After demo]{%
            \includegraphics[width=0.19\linewidth,cframe=orange 1pt]{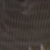}%
        }\hfill%
        \subfloat[Before \& after]{%
            \includegraphics[width=0.19\linewidth,cframe=orange 1pt]{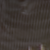}%
        }%
    \end{minipage}%
    \caption{Visual quality comparison with different combinations of iteration sub-blocks before and after demosaicking. Prefiltering and spatio-temporal denoising sub-blocks are used before and after demosaicking with \(\alpha=0.5\). \reviewed{Iterating before and after demosaicking clean residual low frequency noise in flat areas (blue patch) without affecting textured areas (orange patch). Scene from the CRVD outdoor dataset.}}%
    \label{fig:ablationiteration}
\end{figure*}

\begin{table}
\centering
\caption{Effect of the iteration sub-blocks. All other sub-blocks are used. Measures taken on MaskdnGAN's test set.}%
\label{tab:ablationiteration}
\begin{tabular}{ccccccc}
  \toprule
  \multirow{2}{*}[-0.5\dimexpr\aboverulesep+\belowrulesep+\cmidrulewidth]{ISO} & \multicolumn{2}{c}{Iteration} & \multirow{2}{*}[-0.5\dimexpr\aboverulesep+\belowrulesep+\cmidrulewidth]{PSNR} & \multirow{2}{*}[-0.5\dimexpr\aboverulesep+\belowrulesep+\cmidrulewidth]{SSIM} & \multirow{2}{*}[-0.5\dimexpr\aboverulesep+\belowrulesep+\cmidrulewidth]{MS-SSIM} & \multirow{2}{*}[-0.5\dimexpr\aboverulesep+\belowrulesep+\cmidrulewidth]{VMAF}   \\
  \cmidrule(lr){2-3}
  & RAW & Demo & & & & \\
  \midrule
  \multirow{4}{*}{1600} & \xmark & \xmark & 40.64 & 0.9975 & 0.9955 & 98.90 \\ 
                        & \xmark & \cmark & 40.66 & \textbf{0.9976} & 0.9956 & 98.92 \\ 
                        & \cmark & \xmark & \textbf{40.75} & \textbf{0.9976} & 0.9956 & 98.96 \\ 
                        & \cmark & \cmark & 40.74 & \textbf{0.9976} & \textbf{0.9957} & \textbf{98.98} \\ 
  \midrule
  \multirow{4}{*}{3200} & \xmark & \xmark & 38.81 & \textbf{0.9961} & 0.9930 & 98.31 \\ 
                        & \xmark & \cmark & 38.86 & \textbf{0.9961} & 0.9931 & 98.34 \\ 
                        & \cmark & \xmark & 38.96 & \textbf{0.9961} & 0.9932 & 98.40 \\ 
                        & \cmark & \cmark & \textbf{38.97} & \textbf{0.9961} & \textbf{0.9933} & \textbf{98.42} \\ 
  \midrule
  \multirow{4}{*}{6400} & \xmark & \xmark & 37.01 & 0.9935 & 0.9888 & 97.22 \\ 
                        & \xmark & \cmark & 37.08 & \textbf{0.9936} & 0.9891 & 97.31 \\ 
   & \cmark & \xmark & 37.17 & \textbf{0.9936} & 0.9892 & 97.39 \\ 
   & \cmark & \cmark & \textbf{37.21} & \textbf{0.9936} & \textbf{0.9894} & \textbf{97.46} \\ 
  \midrule 
  \multirow{4}{*}{12800} & \xmark & \xmark & 35.06 & 0.9879 & 0.9807 & 93.27 \\ 
   & \xmark & \cmark & 35.16 & 0.9883 & 0.9813 & 93.57 \\ 
   & \cmark & \xmark & 35.19 & 0.9881 & 0.9812 & 93.71 \\ 
   & \cmark & \cmark & \textbf{35.27} & \textbf{0.9885} & \textbf{0.9818} & \textbf{93.95} \\ 
   \bottomrule
\end{tabular}
\end{table}

We also analyze the impact of the iteration sub-blocks by re-enabling them on the scheme.
We keep the prefiltering modules enabled before and after demosaicking.
Fig.~\ref{fig:ablationiteration} shows that these iterations mainly serve the purpose of cleaning up residual, low frequency noise that may remain in flat zones.
Textured zones (like the orange patch) are visually unaffected.
Table~\ref{tab:ablationiteration} shows that iteration sub-blocks slightly improve results in all cases, most noticeably at higher ISOs.

Finally, using the full unrestricted pipeline, we enlarge the temporal window size \(W\) from \(3\) (\(T_{b} = T_{f} = 1\)) to \(5\) (\(T_{b} = T_{f} = 2\)) to measure its influence on the quality of the results.
Table~\ref{tab:ablationwindow} shows that a larger window size only improves the results marginally when the noise level is low.
However, as it increases, the improvement becomes more significant, seeing as the extended temporal information becomes crucial to reconstruct obfuscated details in regions with low signal to noise ratios.

\begin{table}
\centering
\caption{Effect of the temporal window size. All sub-blocks are used. Measures taken on MaskdnGAN's test set.}%
\label{tab:ablationwindow}
\begin{tabular}{cccccc}
  \toprule
  ISO & \(W\) & PSNR & SSIM & MS-SSIM & VMAF \\
  \midrule
  \multirow{2}{*}{1600} & 3 & 40.74 & 0.9976 & 0.9957 & 98.98 \\ 
                        & 5 & \textbf{40.76} & \textbf{0.9977} & \textbf{0.9958} & \textbf{99.03} \\ 
  \midrule
  \multirow{2}{*}{3200} & 3 & 38.97 & 0.9961 & 0.9933 & 98.42 \\ 
                        & 5 & \textbf{39.11} & \textbf{0.9964} & \textbf{0.9938} & \textbf{98.55} \\ 
  \midrule
  \multirow{2}{*}{6400} & 3 & 37.21 & 0.9936 & 0.9894 & 97.46 \\ 
                        & 5 & \textbf{37.48} & \textbf{0.9943} & \textbf{0.9904} & \textbf{97.73} \\ 
  \midrule 
  \multirow{2}{*}{12800} & 3 & 35.27 & 0.9885 & 0.9818 & 93.95 \\ 
                         & 5 & \textbf{35.66} & \textbf{0.9898} & \textbf{0.9840} & \textbf{95.13} \\ 
   \bottomrule
\end{tabular}
\end{table}

\subsection{Comparison with the State of the Art}

To demonstrate the competitiveness of our method, we compare it with several recent state-of-the-art neural networks for RAW video denoising whose \reviewed{official implementations are} available online.
In particular, we compare it with RViDenet~\cite{Yue2020RViDeNet}, MaskDnGAN~\cite{Paliwal2021MaskDnGAN} and UDVD~\cite{Sheth2021UDVD}, fixing \(T_{b}=T_{f}=2\). 
\reviewed{We also measure it against two JDD schemes: the classical FlexISP~\cite{Heide2014FlexISP} and the newer deep learning-based video algorithm RVDD~\cite{Dewil2023RVDD}.}
\reviewed{Since, to the best of our knowledge, no classical JDD methods have been developed for video, we settle for a frame-by-frame application of FlexISP.}
We show that the nature of our method makes it more adaptable to changes in the noise level (i.e.\ the ISO value) and \reviewed{database}, which is currently a challenge for deep learning when applied to real photography. 

All the networks we compare with have been trained on sequences with CRVD-like noise.
Since the datasets we use have noise realizations from this same noise distribution, we do not retrain the networks and just use the pretrained models the authors provide.
\reviewed{In the case of RVDD, where only checkpoints trained on ISOs 3200 and 12800 are supplied, we share the results for the best working one, which always happens to be the one trained with an ISO immediately higher than that of the noisy data.}
To guarantee fair visual and quantitative comparisons, we apply exactly the same ISP to all methods, with the exception that we white balance the images after denoising to respect the networks' expected inputs.

\begin{table}
\centering
\caption{Quantitative comparisons on MaskDnGAN's test set. All measures are taken on the RGB images after applying the full ISP. The applied ISP is the same for all methods. }
\label{tab:sotamaskdngan}
\begin{tabular}{clcccc}
  \toprule
ISO & Method & PSNR & SSIM & MS-SSIM & VMAF \\ 
  \midrule
   \multirow{4}{*}{1600} & FlexISP~\cite{Heide2014FlexISP} & 32.52 & 0.9900 & 0.9839 & 85.82 \\ 
   & RViDeNet~\cite{Yue2020RViDeNet} & 38.88 & 0.9962 & 0.9931 & 98.48 \\ 
   & UDVD~\cite{Sheth2021UDVD} & 35.28 & 0.9942 & 0.9880 & 93.91 \\ 
   & MaskDnGAN~\cite{Paliwal2021MaskDnGAN} & \textbf{40.93} & \underline{0.9976} & \underline{0.9957} & \underline{99.06} \\ 
   & RVDD~\cite{Dewil2023RVDD} & 39.15 & 0.9968 & 0.9949 & \textbf{99.23} \\
   & Ours & \underline{40.76} & \textbf{0.9977} & \textbf{0.9958} & 99.03 \\ 
\midrule
   \multirow{4}{*}{3200} & FlexISP~\cite{Heide2014FlexISP} & 31.28 & 0.9840 & 0.9739 & 91.47 \\ 
   & RViDeNet~\cite{Yue2020RViDeNet} & 37.30 & 0.9937 & 0.9894 & 97.69 \\ 
   & UDVD~\cite{Sheth2021UDVD} & 34.41 & 0.9917 & 0.9838 & 91.42 \\ 
   & MaskDnGAN~\cite{Paliwal2021MaskDnGAN} & \textbf{39.11} & \underline{0.9962} & \underline{0.9934} & \underline{98.59} \\ 
   & RVDD~\cite{Dewil2023RVDD} & 37.49 & 0.9955 & 0.9919 & \textbf{98.84} \\
   & Ours & \textbf{39.11} & \textbf{0.9964} & \textbf{0.9938} & 98.55 \\ 
\midrule
   \multirow{4}{*}{6400} & FlexISP~\cite{Heide2014FlexISP} & 29.76 & 0.9743 & 0.9580 & 82.83 \\ 
   & RViDeNet~\cite{Yue2020RViDeNet} & 35.55 & 0.9895 & 0.9831 & 95.11 \\ 
   & UDVD~\cite{Sheth2021UDVD} & 33.38 & 0.9875 & 0.9774 & 87.21 \\ 
   & MaskDnGAN~\cite{Paliwal2021MaskDnGAN} & \underline{37.38} & \underline{0.9942} & \underline{0.9900} & \underline{97.86} \\ 
   & RVDD~\cite{Dewil2023RVDD} & 36.98 & 0.9931 & 0.9899 & \textbf{98.05} \\
   & Ours & \textbf{37.48} & \textbf{0.9943} & \textbf{0.9904} & 97.73 \\ 
\midrule
   \multirow{4}{*}{12800} & FlexISP~\cite{Heide2014FlexISP} & 27.98 & 0.9579 & 0.9324 & 72.17 \\ 
   & RViDeNet~\cite{Yue2020RViDeNet} & 33.64 & 0.9818 & 0.9716 & 87.86 \\ 
   & UDVD~\cite{Sheth2021UDVD} & 32.18 & 0.9804 & 0.9672 & 80.99 \\ 
   & MaskDnGAN~\cite{Paliwal2021MaskDnGAN} & \underline{35.65} & \textbf{0.9907} & \textbf{0.9847} & \underline{95.52} \\ 
   & RVDD~\cite{Dewil2023RVDD} & 34.91 & 0.9894 & 0.9822 & \textbf{96.25} \\
   & Ours & \textbf{35.66} & \underline{0.9898} & \underline{0.9840} & 95.13 \\ 
   \bottomrule
\end{tabular}
\end{table}

\begin{figure*}
    \centering
    \subfloat[Scene 3 (ISO 3200)]{%
        \includegraphics[width=0.245\linewidth]{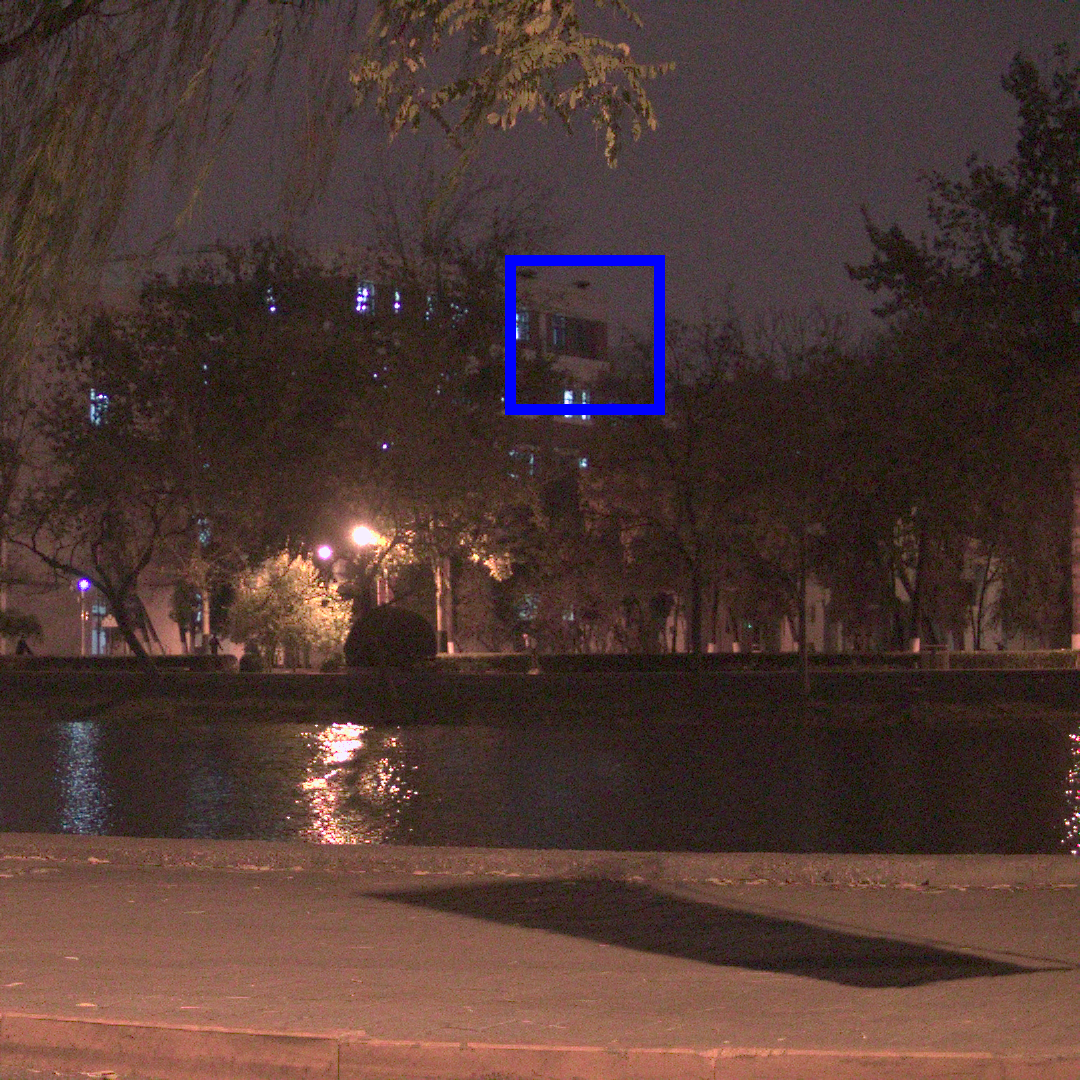}%
    }\hfill%
    \subfloat[Noisy]{%
        \includegraphics[width=0.245\linewidth,cframe=blue 1pt]{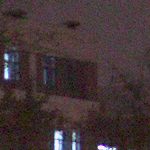}%
    }\hfill%
    \subfloat[FlexISP]{%
        \includegraphics[width=0.245\linewidth,cframe=blue 1pt]{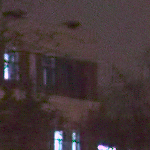}%
    }\hfill%
    \subfloat[RViDeNet]{%
        \includegraphics[width=0.245\linewidth,cframe=blue 1pt]{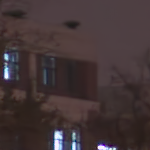}%
    }\par\smallskip%
    \subfloat[UDVD]{%
        \includegraphics[width=0.245\linewidth,cframe=blue 1pt]{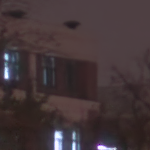}%
    }\hfill%
    \subfloat[RVDD]{%
        \includegraphics[width=0.245\linewidth,cframe=blue 1pt]{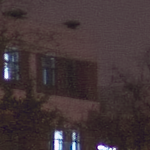}%
    }\hfill%
    \subfloat[MaskDnGAN]{%
        \includegraphics[width=0.245\linewidth,cframe=blue 1pt]{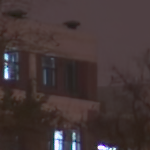}%
    }\hfill%
    \subfloat[Ours]{%
        \includegraphics[width=0.245\linewidth,cframe=blue 1pt]{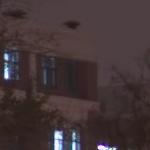}%
    }%
    \caption{\reviewed{Comparison against state-of-the-art video denoising methods on a real outdoor scene with low light from the CRVD dataset with ISO 3200.}}%
    \label{fig:sotacrvd1}
\end{figure*}

\begin{figure*}
    \centering
    \subfloat[Scene 6 (ISO 12800)]{%
        \includegraphics[width=0.245\linewidth]{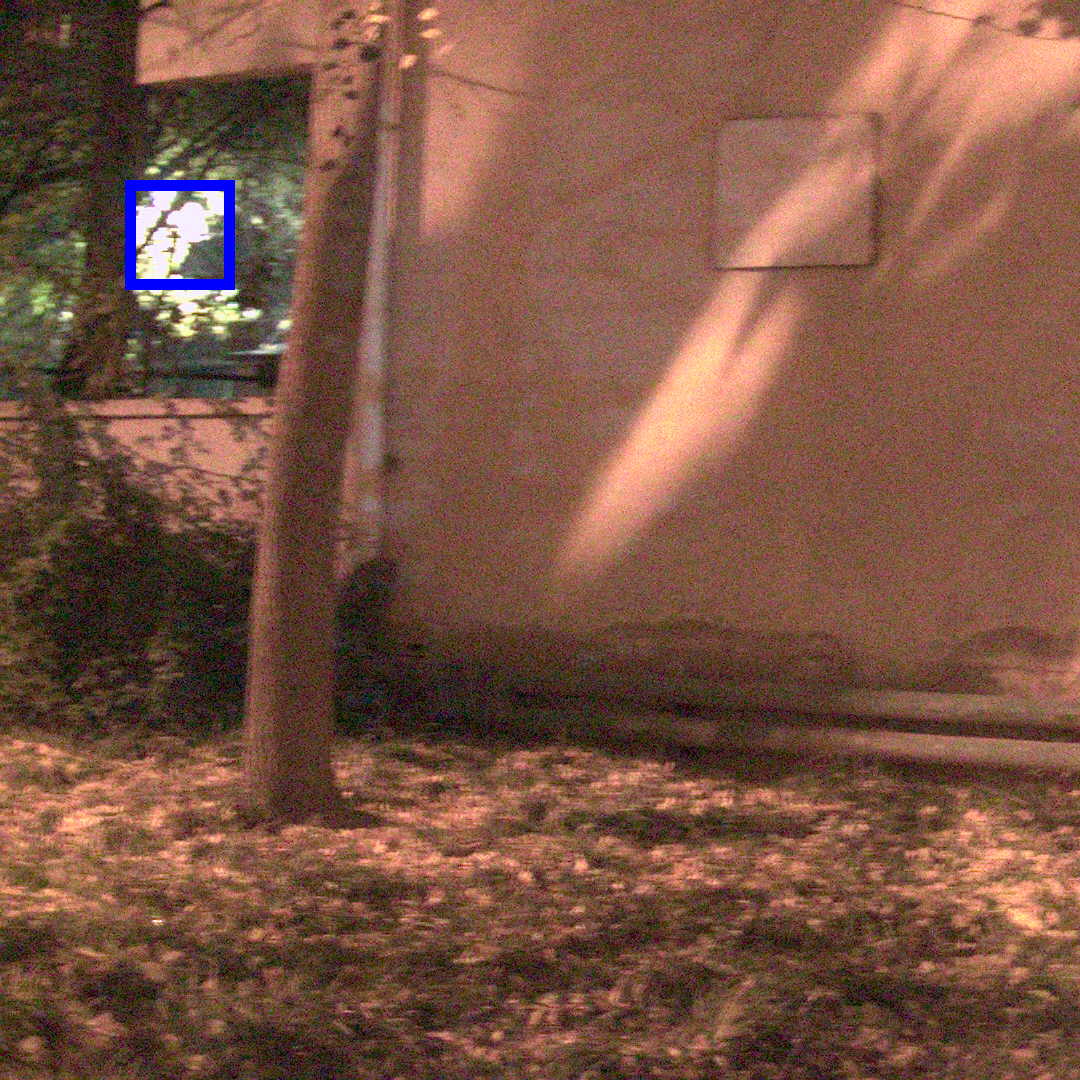}%
    }\hfill%
    \subfloat[Noisy]{%
        \includegraphics[width=0.245\linewidth,cframe=blue 1pt]{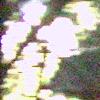}%
    }\hfill%
    \subfloat[FlexISP]{%
        \includegraphics[width=0.245\linewidth,cframe=blue 1pt]{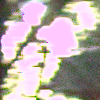}%
    }\hfill%
    \subfloat[RViDeNet]{%
        \includegraphics[width=0.245\linewidth,cframe=blue 1pt]{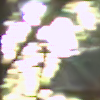}%
    }\par\smallskip%
    \subfloat[UDVD]{%
        \includegraphics[width=0.245\linewidth,cframe=blue 1pt]{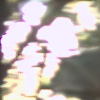}%
    }\hfill%
    \subfloat[RVDD]{%
        \includegraphics[width=0.245\linewidth,cframe=blue 1pt]{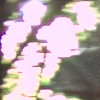}%
    }\hfill%
    \subfloat[MaskDnGAN]{%
        \includegraphics[width=0.245\linewidth,cframe=blue 1pt]{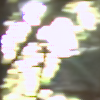}%
    }\hfill%
    \subfloat[Ours]{%
        \includegraphics[width=0.245\linewidth,cframe=blue 1pt]{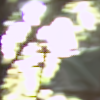}%
    }%
    \caption{\reviewed{Comparison against state-of-the-art video denoising methods on a real outdoor scene with locally strong illumination from the CRVD dataset with ISO 12800. Our method correctly recovers the colors and textures in and around the sunbeam, while other methods leave artifacts due to saturation.}}%
    \label{fig:sotacrvd2}
\end{figure*}

\subsubsection{Numerical Comparison}

Table~\ref{tab:sotamaskdngan} summarizes the quantitative results of the different methods on MaskDnGAN's synthetic test set.
Our method outperforms \reviewed{FlexISP}, RViDeNet and UDVD in all cases \reviewed{and metrics}.
\reviewed{The same occurs against RVDD in PSNR, SSIM and MS-SSIM but not VMAF, where RVDD stands out above the rest.}  
\reviewed{Our method also} surpasses or matches MaskDnGAN in PSNR for ISOs ranging from 3200 to 12800, in SSIM and MS-SSIM for the range 1600 to 6400, and keeps close behind it in VMAF\@.
\reviewed{It must be noted that this dataset was used for the validation of MaskDnGAN, so the network has been optimized specifically for it.}

We switch our experiments to ARRI~\cite{Andriani2013ARRI} \reviewed{with different ISO values than the previous dataset}.
We fit a linear model to \reviewed{both of} the noise calibration parameters \(a\) and \(\sqrt{b}\) provided by CRVD---which approximately double when doubling the ISO---, and use it to add noise realizations with ISOs 777, 1989, \reviewed{3200} and 4869 to the sequences.
Table~\ref{tab:sotaarri} shows the results on these conditions.
\reviewed{Our method obtains better PSNR, SSIM and MS-SSIM values than all the other methods for all ISOs, being a close second in VMAF after RVDD\@.}
\reviewed{In PSNR, concretely, the difference between our method and the second best ranges from \SI{1}{\dB} to almost \SI{2}{\dB}}.
\reviewed{We note that RVDD still attains high values in all metrics in ISOs 777 and 1989 even when trained with ISO 3200.}
\reviewed{However, at ISO 4869 using the ISO 12800 checkpoint (since the ISO 3200 one is not able to completely remove noise), it is overtaken by UDVD in all metrics except VMAF.}
\reviewed{This highlights the lack of adaptability of RVDD to wide-ranging noise levels as its main drawback.}
\reviewed{As opposed to the previous dataset}, UDVD \reviewed{also} performs better than MaskDnGAN \reviewed{in PSNR, SSIM and MS-SSIM} for these images.
\reviewed{{Indeed, the numbers for MaskDnGAN agree with the discussion by its authors specifically citing its generalization capability as its main limitation.}}

\subsubsection{Visual Comparison}

\reviewed{Fig.~\ref{fig:sotacrvd1} shows some results for the different methods on a real low-light sequence from the outdoor CRVD dataset.}
Our method is able to recover complex structures and texture while removing all noise.
\reviewed{FlexISP is not able to remove it completely and leaves some artifacts that are noticeable in regions with low signal-to-noise ratios.}
RViDeNet smoothes some textures too much in comparison, particularly in regions with fine details.
UDVD can present reconstruction artifacts that are discernible upon zooming in close to the image, while also oversmoothing textures.
\reviewed{RVDD stands out in its capacity to reconstruct fine details, but at the cost of leaving residual noise that is especially perceptible in flat regions.}
\reviewed{MaskDnGAN, being the best of the rest of methods according to Table~\ref{tab:sotamaskdngan}, still blurs some texture.}

\reviewed{In Fig.~\ref{fig:sotacrvd2}, we compare the application of the different methods on a patch with locally strong illumination.}
\reviewed{FlexISP and RVDD exhibit a color shift towards red in the saturated region.}
\reviewed{RViDeNet, UDVD and MaskDnGAN are not able to remove the small pixel artifacts inside and at the edges of the saturated region.}
\reviewed{Our method is able to deal with these issues.}

\begin{table}
\centering
\caption{Quantitative comparisons on the ARRI dataset. All measures are taken on the RGB images after applying the full ISP. The applied ISP is the same for all methods.}%
\label{tab:sotaarri}
\begin{tabular}{clcccc}
  \toprule
ISO & Method & PSNR & SSIM & MS-SSIM & VMAF \\ 
  \midrule
   \multirow{4}{*}{777} & FlexISP~\cite{Heide2014FlexISP} & 35.61 & 0.9937 & 0.9869 & 86.07 \\ 
                        & RViDeNet~\cite{Yue2020RViDeNet} & 36.26 & 0.9939 & 0.9812 & 81.15 \\ 
                        & UDVD~\cite{Sheth2021UDVD} & 38.46 & 0.9970 & 0.9904 & 85.14 \\ 
                        & MaskDnGAN~\cite{Paliwal2021MaskDnGAN} & 38.48 & 0.9962 & 0.9887 & 87.97 \\ 
                        & RVDD~\cite{Dewil2023RVDD} & \underline{39.05} & \underline{0.9973} & \underline{0.9927} & \textbf{94.49} \\
                        & Ours & \textbf{40.92} & \textbf{0.9977} & \textbf{0.9946} & \underline{92.97} \\ 
   \midrule
   \multirow{4}{*}{1989} & FlexISP~\cite{Heide2014FlexISP} & 34.10 & 0.9899 & 0.9771 & 80.34 \\ 
                         & RViDeNet~\cite{Yue2020RViDeNet} & 35.86 & 0.9930 & 0.9791 & 79.82 \\ 
                         & UDVD~\cite{Sheth2021UDVD} & 37.96 & 0.9962 & 0.9881 & 83.64 \\ 
                         & MaskDnGAN~\cite{Paliwal2021MaskDnGAN} & 37.60 & 0.9953 & 0.9866 & 86.37 \\ 
                         & RVDD~\cite{Dewil2023RVDD} & \underline{38.79} & \underline{0.9969} & \underline{0.9921} & \textbf{92.78} \\
                         & Ours & \textbf{39.96} & \textbf{0.9970} & \textbf{0.9928} & \underline{91.51} \\ 
   \midrule
   \multirow{4}{*}{3200} & FlexISP~\cite{Heide2014FlexISP} & 33.07 & 0.9866 & 0.9685 & 76.64 \\ 
                         & RViDeNet~\cite{Yue2020RViDeNet} & 35.66 & 0.9922 & 0.9773 & 78.75 \\ 
                         & UDVD~\cite{Sheth2021UDVD} & 37.59 & 0.9955 & 0.9863 & 82.35 \\ 
                         & MaskDnGAN~\cite{Paliwal2021MaskDnGAN} & 37.18 & 0.9947 & 0.9849 & 84.89 \\ 
                         & RVDD~\cite{Dewil2023RVDD} & \underline{37.60} & \underline{0.9960} & \underline{0.9882} & \textbf{90.58} \\
                         & Ours & \textbf{39.40} & \textbf{0.9965} & \textbf{0.9913} & \underline{90.45} \\ 
   \midrule
   \multirow{4}{*}{4869} & FlexISP~\cite{Heide2014FlexISP} & 32.03 & 0.9826 & 0.9581 & 72.61 \\ 
                         & RViDeNet~\cite{Yue2020RViDeNet} & 34.90 & 0.9905 & 0.9734 & 75.74 \\ 
                         & UDVD~\cite{Sheth2021UDVD} & \underline{37.22} & \underline{0.9947} & \underline{0.9843} & 80.80 \\ 
                         & MaskDnGAN~\cite{Paliwal2021MaskDnGAN} & 36.21 & 0.9933 & 0.9817 & 82.35 \\ 
                         & RVDD~\cite{Dewil2023RVDD} & 36.72 & 0.9943 & 0.9865 & \textbf{91.57} \\
                         & Ours & \textbf{38.87} & \textbf{0.9958} & \textbf{0.9896} & \underline{89.06} \\ 
   \bottomrule
\end{tabular}
\end{table}

\begin{figure*}
    \centering
    \subfloat[ARRI Lake (ISO 1989)]{%
        \includegraphics[width=0.245\linewidth]{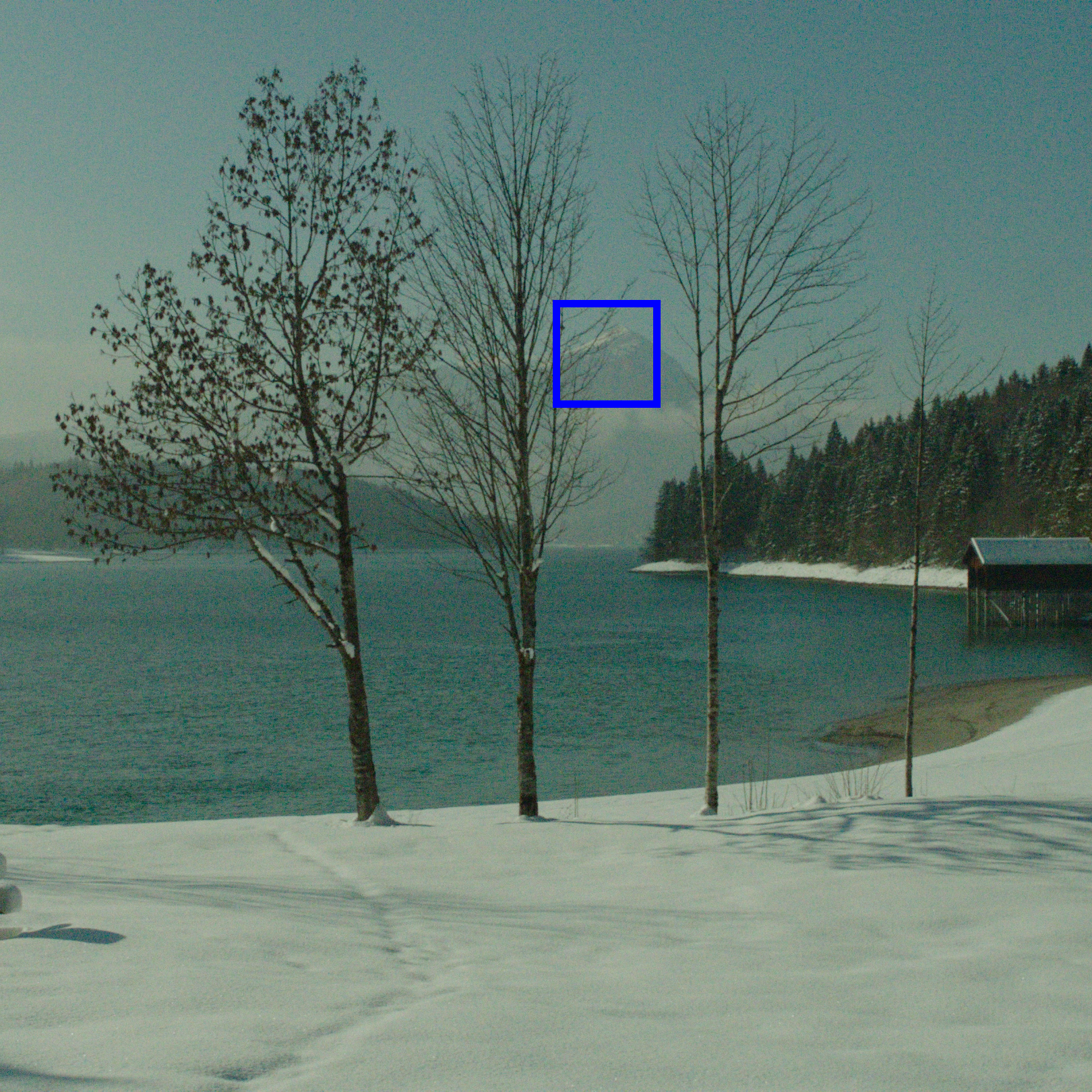}%
    }\hfill%
    \subfloat[Noisy / GT]{%
        \includegraphics[width=0.245\linewidth,cframe=blue 1pt]{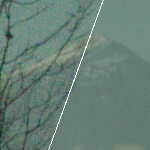}%
    }\hfill%
    \subfloat[FlexISP]{%
        \includegraphics[width=0.245\linewidth,cframe=blue 1pt]{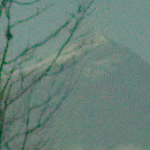}%
    }\hfill%
    \subfloat[RViDeNet]{%
        \includegraphics[width=0.245\linewidth,cframe=blue 1pt]{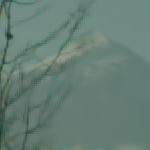}%
    }\par\smallskip%
    \subfloat[UDVD]{%
        \includegraphics[width=0.245\linewidth,cframe=blue 1pt]{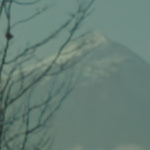}%
    }\hfill%
    \subfloat[MaskDnGAN]{%
        \includegraphics[width=0.245\linewidth,cframe=blue 1pt]{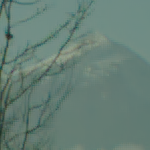}%
    }\hfill%
    \subfloat[RVDD]{%
        \includegraphics[width=0.245\linewidth,cframe=blue 1pt]{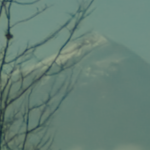}%
    }\hfill%
    \subfloat[Ours]{%
        \includegraphics[width=0.245\linewidth,cframe=blue 1pt]{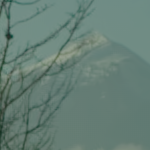}%
    }%
    \caption{\reviewed{Comparison against state-of-the-art video denoising methods on the real scene ARRI Lake with ISO 1989.}}%
    \label{fig:sotaarri1}
\end{figure*}

\begin{figure*}
    \centering
    \subfloat[ARRI Pool (ISO 4869)]{%
        \includegraphics[width=0.245\linewidth]{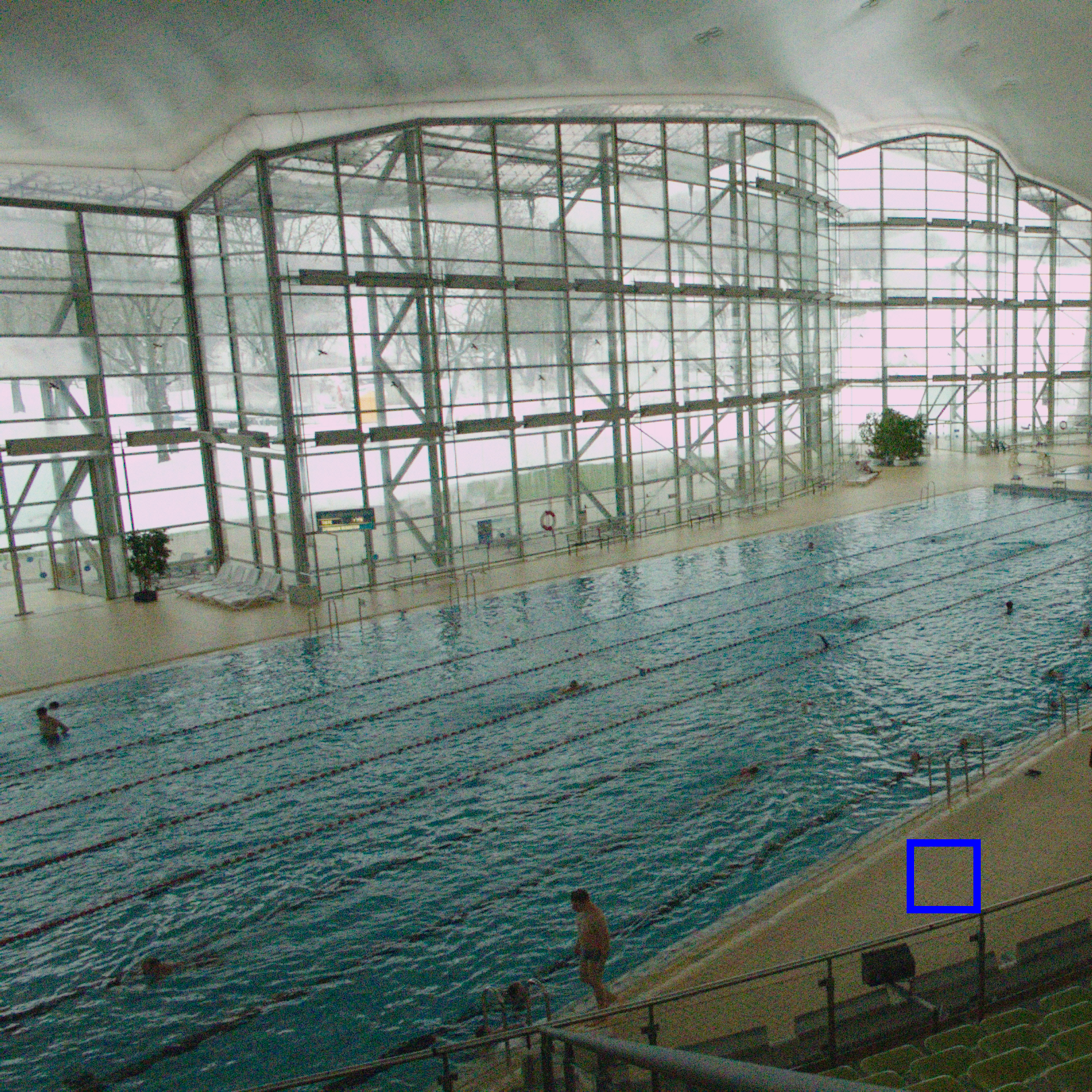}%
    }\hfill%
    \subfloat[Noisy / GT]{%
        \includegraphics[width=0.245\linewidth,cframe=blue 1pt]{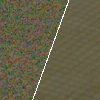}%
    }\hfill%
    \subfloat[FlexISP]{%
        \includegraphics[width=0.245\linewidth,cframe=blue 1pt]{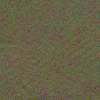}%
    }\hfill%
    \subfloat[RViDeNet]{%
        \includegraphics[width=0.245\linewidth,cframe=blue 1pt]{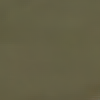}%
    }\par\smallskip%
    \subfloat[UDVD]{%
        \includegraphics[width=0.245\linewidth,cframe=blue 1pt]{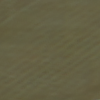}%
    }\hfill%
    \subfloat[MaskDnGAN]{%
        \includegraphics[width=0.245\linewidth,cframe=blue 1pt]{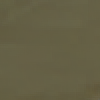}%
    }\hfill%
    \subfloat[RVDD]{%
        \includegraphics[width=0.245\linewidth,cframe=blue 1pt]{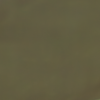}%
    }\hfill%
    \subfloat[Ours]{%
        \includegraphics[width=0.245\linewidth,cframe=blue 1pt]{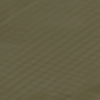}%
    }%
    \caption{\reviewed{Comparison against state-of-the-art video denoising methods on the real scene ARRI Pool with ISO 4869.}}%
    \label{fig:sotaarri2}
\end{figure*}

Fig.~\ref{fig:sotaarri1} and  \ref{fig:sotaarri2} show examples on two sequences and ISOs of the ARRI dataset.
\reviewed{Again, FlexISP does not remove the noise completely.}
RViDeNet and MaskDnGAN have trouble reconstructing textured areas and present very noticeable artifacts on edges.
UDVD does a better job with edges and complex textures, but still has artifacts similar to those in Fig.~\ref{fig:sotacrvd2}.
\reviewed{RVDD removes noise and recovers fine details correctly in Fig.~\ref{fig:sotaarri1} (ISO 1989), but oversmoothes textures in Fig.~\ref{fig:sotaarri2} (ISO 4869).}
Our method, on the other hand, is able to adapt correctly to the noise conditions and returns visually pleasant results.

\subsubsection{Computational Resources}

\begin{table}
    \centering
    \caption{Computational resources used per frame when denoising a sequence with images of size \(256 \times 256\) pixels.}%
    \label{tab:computationalresources}
    \begin{tabular}{crc}
        \toprule
        Method & Time (s) & Peak memory usage (GB) \\
		\midrule
        FlexISP & 78.19 & 0.824 \\
        RViDeNet & 15.45 & 2.965 \\
        UDVD & 15.30 & 3.084 \\
        MaskDnGAN & 20.96 & 2.986 \\
        RVDD & 3.46 & 3.649 \\
        Ours & 23.58 & 0.214 \\
		\bottomrule
	\end{tabular}
\end{table}

\reviewed{To assess the computational resources that our method uses, we denoise a test sequence consisting of 10 frames of size \(256 \times 256\) pixels.}
\reviewed{In Table~\ref{tab:computationalresources} we share the mean ISP time per frame (that is, the time it takes to obtain the final sRGB image from a CFA) and the peak memory usage of all methods.}
\reviewed{Bear in mind that the showcased times are dependent on many factors, and are not illustrative of the applicability of these methods to real scenarios such as mobile videography.}
\reviewed{Case in point, the C++ implementation of our method and the MATLAB implementation of FlexISP run on the CPU, while the Python implementation of the neural networks are made to be run in a GPU.}
\reviewed{Other considerations such as multiprocessing, I/O operations, and optimization of the code are also not taken into account.}
\reviewed{For a realistic comparison, an efficient onboard implementation of the methods on a mobile phone would be required, which is currently not viable for the aforementioned networks due to memory constraints.}
\reviewed{Table~\ref{tab:computationalresources} shows that at least around \SI{3}{\giga\byte} of memory are needed to employ these networks.}
\reviewed{Our method uses around 14 times less memory compared to them, and almost 4 times compared to FlexISP, which makes a mobile implementation feasible.}


\section{Conclusion}

We have presented a self-similarity-based video denoising scheme that exploits both the data at the camera sensor and information on the RGB domain.
We have shown that a correct balance between two denoisers---one before and one after demosaicking---significantly improves image quality over denoising only before or after demosaicking. The optimal balance depends on the noise level, increasing the RAW filtering for higher ISO values.

We have also integrated a temporal trajectory prefiltering step before the usual spatio-temporal denoising stage
and studied its effects on texture reconstruction.
We have found that, with strong restrictions on the motion model to avoid misalignments, this step can greatly help in the recovery of fine details, especially at high noise levels.

The proposed method only requires an estimation of the noise model at the sensor, which can be obtained easily with a simple camera calibration procedure.
This allows an accurate noise removal at any noise level, making our method suitable for real-world videography.
Experiments show the competitiveness of our approach against state-of-the-art deep learning-based methods in terms of visual quality, and demonstrate that it can better adapt to general video sequences.

\bibliographystyle{IEEEtran}
\bibliography{references}

\end{document}